\documentclass[]{aastex631}
\usepackage{amsmath}
\usepackage{mathrsfs}
\usepackage{CJK}
\usepackage{threeparttable}
\usepackage{captcont}
\usepackage{xcolor}
\usepackage{booktabs}
\usepackage{soul}
\usepackage{multirow}

\shorttitle{{Exploring Formation Channels}}
\shortauthors{Li et al.}

\graphicspath{{./}{figures/}}

\begin{document}
\begin{CJK*}{UTF8}{gbsn}

\title{Exploring Field-evolution and Dynamical-capture Coalescing Binary Black Holes in GWTC-3}

\author[0000-0001-5087-9613]{Yin-Jie Li（李银杰）}
\affiliation{Key Laboratory of Dark Matter and Space Astronomy, Purple Mountain Observatory, Chinese Academy of Sciences, Nanjing 210023, People's Republic of China}

\author[0000-0001-9120-7733]{Shao-Peng Tang（唐少鹏）}
\affiliation{Key Laboratory of Dark Matter and Space Astronomy, Purple Mountain Observatory, Chinese Academy of Sciences, Nanjing 210023, People's Republic of China}

\author[0000-0002-0822-0337]{Shi-Jie Gao (高世杰)}
\affiliation{School of Astronomy and Space Science, Nanjing University, Nanjing, 210023, People’s Republic of China}
\affiliation{Key Laboratory of Modern Astronomy and Astrophysics, Nanjing University, Ministry of Education, Nanjing, 210023, People’s Republic of China}

\author{Dao-Cheng Wu (伍道成)}
\affiliation{Department of mathematics, University of Michigan, Ann Arbor, 48109, United State of America}
\affiliation{School of Mathematics and Physics, Xi'an Jiaotong-liverpool University, Suzhou, 215123, People's Republic of China}

\author[0000-0001-9626-9319]{Yuan-Zhu Wang（王远瞩）}
\affiliation{Institute for theoretical physics and cosmology, Zhejiang University of Technology, Hangzhou, 310032, People's Republic of China}
\affiliation{Key Laboratory of Dark Matter and Space Astronomy, Purple Mountain Observatory, Chinese Academy of Sciences, Nanjing 210023, People's Republic of China}
\email{The corresponding author: vamdrew@zjut.edu.cn (Y.Z.W)}

\begin{abstract}
We investigate formation channels for merging binary black holes (BBHs) in GWTC-3, with a dedicated semiparametric population model. The model first describes or excludes a high-spin (with magnitudes of $\sim0.7$) and high-mass (ranging in $\sim 20-80M_{\odot}$) subpopulation, which was identified by previous works and can be interpreted as hierarchical mergers. We find that the rest of BBH population can be categorized into two subpopulations with different mass and mass-ratio distributions, as indicated by a Bayes factor of $\ln\mathcal{B}=1.8$. One subpopulation, characterized by nearly aligned spins and consistent with isolated-field formation, likely dominates the 10-solar-mass peak in the primary-mass function. The other subpopulation, with isotropic spins and consistent with the dynamical channels, shows a stronger preference for symmetric pairing, and mainly contributes to the 35-solar-mass peak in the primary-mass function. Note that the Bayes factor is not high enough with the currently available data, so that the case of a single population is still acceptable. Additionally, we compare the mass distributions between merging BBHs and black holes (BHs) in high-mass X-ray binaries (HMXBs).  We find that the primary mass of the aligned subpopulation is slightly lighter than those of the HMXB BHs, while the isotropic subpopulation is consistent with the HMXB BHs, if the power-law index of its mass function is shifted by 2, as indicated by the dynamical formation channels. However, the spin magnitudes of both subpopulations are significantly smaller than those of the HMXB BHs.

\end{abstract} 


\section{Introduction}

Gravitational waves (GWs) can { provide us with} information about coalescing compact binaries \citep{2016PhRvL.116f1102A}, including component masses, spin magnitudes, spin orientations, and luminosity distances of the sources. { With these inferred properties, we can investigate the formation and evolution of the coalescing compact binaries. Various formation channels have been proposed \citep[e.g.,][]{1988ApJ...329..764L,2016A&A...588A..50M,2016MNRAS.458.2634M,2018PhRvL.121p1103F,2019PhRvL.123r1101Y}. Generally, formation channels can be divided into two categories: isolated field binary evolution and dynamical formations \citep[see][and their references.]{2018arXiv180909130M,2022PhR...955....1M}.}
With the rapidly increasing GW detections of compact binary coalescences (CBCs) \citep{2019PhRvX...9c1040A,2021PhRvX..11b1053A,2021arXiv210801045T,2023PhRvX..13d1039A}, including binary neutron stars (BNSs) \citep{2017PhRvL.119p1101A,2020ApJ...892L...3A}, binary black holes (BBHs) \citep{2016PhRvL.116f1102A}, and neutron star-black holes (NSBHs) \citep{2021ApJ...915L...5A}, we can reveal the formation and evolution of the CBCs { through statistical analysis} \citep[e.g.,][]{2020ApJ...899L...8F,2020ApJ...891L..27F,2021ApJ...913...42W,2021ApJ...922....3T,2021ApJ...923...97L,2023PhRvX..13a1048A}. { Different formation scenarios result in CBCs with distinct distributions of physical parameters such as mass, mass ratio, spin magnitude, and spin orientation.} {For instance, hierarchical mergers exhibit typical spin magnitudes ($\sim0.7$) \citep{2017ApJ...840L..24F,2017PhRvD..95l4046G} and unusually high masses (lying in the pair-instability mass-gap \citep{2017ApJ...836..244W,2021ApJ...912L..31W}), which makes them distinguishable \citep{2021NatAs...5..749G}.} 

The isolated field evolution and the dynamical formation channels are expected to be distinguished by the spin-orientation distributions of the BBHs \citep{2022PhR...955....1M}. BBHs from isolated-field evolution are expected to have spins nearly aligned with the orbital angular momentum \citep{2016ApJ...832L...2R,2018PhRvD..98h4036G}. In contrast, dynamically formed BBHs in star clusters are expected to have isotropic spins \citep{2017MNRAS.471.2801S,2018ApJ...854L...9F}. Note that dynamical formation channels in gas-rich environments, such as the disks of active galactic nuclei (AGNs), may also produce BBHs with aligned spins \citep{2019PhRvL.123r1101Y,2021MNRAS.507.3362T}.  \citet{2023PhRvX..13a1048A} have investigated the spin-orientation ($\cos\theta_{1,2}$, i.e., cosine of the tilt angle between component spin and a binary's orbital angular momentum) distribution of the BBHs using a parametric model $\pi(\cos\theta_{1,2}|\sigma_{\rm t},\zeta)=\zeta \mathcal{G}(\cos\theta_{1,2}|1,\sigma_{\rm t},-1,1)+(1-\zeta)\mathcal{U}(\cos\theta_{1,2}|-1,1)$, so called \textsc{Default} spin model. $\mathcal{U}(\cos\theta_{1,2}|-1,1)$ is a Uniform distribution in (-1,1) and $\mathcal{G}(\cos\theta_{1,2}|1,\sigma_{\rm t},-1,1)$ is a truncated Gaussian distribution in (-1,1), peaking at $\cos\theta_{1,2} = 1$ (perfect alignment) with width of $\sigma_{\rm t}$. They find that either perfectly aligned spins ($\zeta=1$ and $\sigma_{\rm t}\sim0$) or fully isotropic spins ($\zeta=0$ or $\sigma_{\rm t}\gg 1$) are disfavored. 
 However, in their analysis, the $\cos\theta_{1,2}$ distribution is independent from the mass function of the BBHs, which makes it hard to identify the nearly aligned and isotropic-spin subpopulations. {The identification of subpopulations can be benefit from constructing models with mass (or some other parameters) dependent spin distributions, as these models can account for the potential parameter correlations among in the BBH populations \citep[e.g.,][]{2022ApJ...941L..39W,2024arXiv240603257G}.}

\cite{2024PhRvL.133e1401L} was the first to use a flexible mixture model to search for the subpopulations of coalescing BHs, identifying two subpopulations consistent with first-generation / stellar-formed and higher-generation / merger-formed BHs. These findings were recently confirmed by another study \citep{2024arXiv240601679P}. \cite{2023arXiv230401288G} {has} also searched for the subpopulations of coalescing BBHs using flexible mixture models and found a subpopulation consistent with isolated-field-evolution channels. Although the results of the two works are broadly consistent, the different assumptions in the construction of the mixture model have led to some differing outcomes. For instance, in the model presented by \cite{2024PhRvL.133e1401L}, the spins and masses of two BHs in one BBH are allowed to {follow} different spin-magnitude subpopulations; whereas in the model of \cite{2023arXiv230401288G}, both BHs in a BBH are assumed to follow the same spin distribution. The former is suitable for distinguishing hierarchical mergers \citep{2021NatAs...5..749G}, because higher-generation BHs may be paired {with} first-generation BHs. The latter is more suitable for searching formation channels since both BHs in a BBH must originate from the same formation channel.
\cite{2021NatAs...5..749G} suggests that investigating the occurrence of hierarchical mergers constitutes an orthogonal and complementary direction to the usual ‘field versus dynamics’ formation-channel debate.  
In this work, we search for the subpopulations of field and dynamical formation channels of BBHs, building on the results of \cite{2024PhRvL.133e1401L}.

If the subpopulations of the coalescing BBHs are identified, it would be useful to compare them to the BHs from other observations, such as the XRBs \citep{2006ARA&A..44...49R,2016A&A...587A..61C}.
\citet{2022ApJ...929L..26F} was the first to compare the BHs in XRBs with those in GW sources. They concluded that the two types of BHs are like `Apple and Orange' due to the significant disagreement in their spin-magnitude distributions, although their mass distributions remain consistent. Nevertheless, the GW sources may originate from several channels, and the higher-generation BHs in hierarchical mergers are unlikely to be the same as the BHs in XRBs \citep{2017ApJ...840L..24F,2024PhRvL.133e1401L}. Therefore, it would be more appropriate to compare the BHs from `Apple to Apple', that is, to compare BHs from a single inferred formation channel with those in XRBs. In this study, we focus on comparing GW sources with BHs in high-mass X-ray binaries (HMXBs), as HMXBs are more likely to be the progenitors of merging BBHs due to the presence of massive mass donors.

In this work, we perform hierarchical Bayesian inference to explore the subpopulations of field-evolution and dynamical-capture binary black holes in GWTC-3, building upon previous work \citep{2024PhRvL.133e1401L}, which identified a subpopulation of black holes with spin magnitudes of  $\sim0.7$, consistent with hierarchical mergers.
The remainder of the article is organized as follows: In Section~\ref{sec:meth}, we introduce the inference framework, data, and population models; in Section~\ref{sec:result}, we present the results; and in Section~\ref{sec:check}, we perform model comparisons and mock data studies to validate these results. Finally, we conclude and discuss our findings in Section~\ref{sec:con}.

\section{Methods}\label{sec:meth}
\subsection{Framework and data}
We perform hierarchical Bayesian inferences to infer the population properties of merging compact binaries from GW observations by LIGO/Virgo/KAGRA \citep{2019PhRvX...9c1040A,2021PhRvX..11b1053A,2021arXiv210801045T,2023PhRvX..13d1039A}. The distribution of merging compact binaries is expressed by the population model $\pi(\boldsymbol{\theta_i}|\boldsymbol{\Lambda})$, where $\boldsymbol{\theta_i}$ are the parameters of the individual source including cosmological red-shift $z$, component masses $m_{1,2}$, spin magnitudes $a_{1,2}$, and cosine tilt angles that describe the spin orientations of two components $\cos\theta_{1,2}$; $\boldsymbol{\Lambda}$ are the hyperparameter, which describe the distribution of merging compact binaries. Following \cite{2021ApJ...913L...7A} and \cite{2023PhRvX..13a1048A}, {the likelihood of data $\{\boldsymbol{d}\}$ from $N_{\rm det}$ GW events, given $\boldsymbol{\Lambda}$}, is expressed as 
\begin{equation}\label{eq_llh}
\mathcal{L}(\{\boldsymbol{d}\}, N_{\rm det} |\boldsymbol{\Lambda}, )\propto N^{N_{\rm det}}e^{-N{\xi(\boldsymbol{\Lambda})}}\prod_{i=1}^{N_{\rm det}}\int{\mathcal{L}(\boldsymbol{d_i}|\boldsymbol{\theta_i})\pi(\boldsymbol{\theta_i}|\boldsymbol{\Lambda})d\boldsymbol{\theta_i}},
\end{equation} 
where $N$ is the number of mergers in the Universe over the observation period, which is related to the merger rate.
$\mathcal{L}(\boldsymbol{d_i}|\boldsymbol{\theta_i})$ is the single-event likelihood that can be estimated using the posterior samples (see \cite{2021ApJ...913L...7A} for detail); $\xi(\boldsymbol{\Lambda})$ {is} the detection fraction, and it can be estimated using a Monte Carlo integral over detected injections as introduced in the Appendix {A} of \cite{2021ApJ...913L...7A}.
Since the Monte Carlo summations over samples to approximate the integrals will bring statistical error in the likelihood estimations \cite{2019RNAAS...3...66F,2020arXiv201201317T,2022arXiv220400461E,2022arXiv221012287G,2022ApJ...926...79G,2023MNRAS.526.3495T}, we constrain the prior of hyperparameter to ensure $N_{{\rm eff},i} > 10$\footnote{\citet{2022arXiv220400461E,2022ApJ...937L..13C} suggest that $N_{{\rm eff},i} > 10$ is sufficiently high to ensure accurate marginalization over each event, though \citet{2023PhRvX..13a1048A} adopts a stricter constraint, i.e., $N_{{\rm eff},i} > N_{\rm det}$}, and $N_{\rm eff,sel}> 4 N_{\rm det}$, where $N_{{\rm eff},i}$ and $N_{\rm eff,sel}$ are the effective numbers of samples for $i$-th event and detected injections, respectively, as defined by \cite{2023PhRvX..13a1048A,2019RNAAS...3...66F}. {The injection campaigns can be adopted from (https://zenodo.org/doi/10.5281/zenodo.5636815).}
{In our analysis, we do not account for the spin-induced selection bias when calculating the observable fraction $\xi(\boldsymbol{\Lambda})$, this is because the injection samples are not enough for our model to estimate an accurate $\xi(\boldsymbol{\Lambda})$ \footnote{When accounting for the spin-induced selection effects, the variance of $\xi(\boldsymbol{\Lambda})$ is too large to obtain a reliable inference, and the $N_{\rm eff,sel}$ trend to $4 N_{\rm det}$ which means the injections samples in not enough for an accurate analysis \citep{2019RNAAS...3...66F}. The inferred spin-magnitude distribution strongly peaks at 0.21, a similar circumference is also reported in \cite{2023PhRvD.108j3009G}.}. We find the conclusions in this work are not sensitive to the selection effects, because we have also performed an inference without selection effects, and the classification of the two subpopulations is unchanged, see Figure~\ref{fig:nosel}.}

As for the GW data, we use the `C01:Mixed' posterior samples of BBHs in GWTC-3 \citep{2019PhRvX...9c1040A,2021PhRvX..11b1053A,2021arXiv210801045T,2023PhRvX..13d1039A}, adopted from the Gravitational Wave Open Science Center\footnote{\url{https://www.gw-openscience.org/eventapi/html/GWTC/}}. {We use a size of 5000 for per-event samples instead of the minimum sample size across all events (i.e., 1993 of GW200129\_065458), so that some narrow distributions may also pass the threshold of $N_{{\rm eff},i} > 10$. Note that three events have sample sizes smaller than 5000 (i.e., GW150914\_095045, GW200112\_155838, and GW200129\_065458 have sample sizes of 3337, 4323, and 1993, respectively), for these events, the posterior points are reused by random choice. This manipulation will not increase the effective numbers for these events, but will increase the effective numbers for the events which initially have more than 5000 samples given hyper-parameters $\boldsymbol{\Lambda}$, see {Section~\uppercase\expandafter{\romannumeral 5} in Supplemental Material of \cite{2024PhRvL.133e1401L} for more details}.} Following \cite{2023PhRvX..13a1048A}, we choose a false-alarm rate (FAR) of $1yr^{-1}$ as the threshold to select the events, and exclude GW190814 from our main analysis, since it is a significant outlier \citep{2023PhRvX..13a1048A,2022ApJ...926...34E} in the BBH populations. Consequently, 69 events are selected for our analysis. 
For all the hierarchical inferences, we use the \textit{Pymultinest} \citep{2016ascl.soft06005B} sampler, to obtain the posterior distribution of the hyperparameter. 

\subsection{BBH population models}\label{pop_model}

Previous works \citep{2022ApJ...941L..39W,2024PhRvL.133e1401L} have revealed two subpopulations of coalescing BHs: one has a significantly larger spin-magnitude distribution (peaking at $\sim 0.7$), consistent with the higher-generation BHs in hierarchical mergers \citep{2017ApJ...840L..24F,2017PhRvD..95l4046G,2021NatAs...5..749G}; the other has a smaller spin-magnitude distribution and an upper-mass cutoff (at $\sim 45M_{\odot}$), as expected from pair-instability supernovae \citep{2017ApJ...836..244W,2021ApJ...912L..31W}, consistent with the first-generation (or stellar-formed) BHs. In this work, we aim to {search for} subpopulations mainly rely on the spin-orientation distributions in the low-spin BBHs, which are associated with the first-generation BBHs. 

As frequently studied, the dynamical formation channels predict an isotropic distribution for the spin orientations of BHs \citep{2022PhR...955....1M}; specifically, $\cos\theta_{1,2}\sim \mathcal{U}(-1,1)$. In contrast, isolated-field BBHs are likely to have spin orientations that are nearly aligned with the orbital angular momentum of their systems \citep{2022PhR...955....1M}, where the distribution of $\cos\theta_{1,2}$ can be approximated as $\mathcal{G}(1,\sigma_{\rm ct},-1,1)$. Therefore, we consider the distributions of $\cos\theta_{1,2}$ to be crucial for distinguishing between the field and dynamical formation channels for first-generation (or low-spin) BBHs.
Regarding the component mass functions, we employ the \textsc{PowerLaw Spline} model, as first introduced by \cite{2022ApJ...924..101E}. This flexible mass function is well-suited for uncovering the underlying mass distributions of BBHs from the two formation channels. Such flexible mixture population models (for the mass versus spin distribution) are effective in identifying subpopulations within BBHs, as previously demonstrated by \citet{2024PhRvL.133e1401L} and \citet{2023arXiv230401288G}.

The mass and spin distributions of the {Aligned} (1st) subpopulation can be expressed as 
\begin{equation}\label{eq:onepop}
\pi_{\rm A}({\bf \theta}|{\bf \Lambda}_{\rm A})=P_{\rm m}(m_1,m_2 |{\bf \Lambda}_{\rm A})\times \mathcal{G}(\cos\theta_1,\cos\theta_2|1,\sigma_{\rm t},-1,1)\times\mathcal{G}(a_1,a_2|\mu_{\rm a,1},\sigma_{\rm a,1},0,1),
\end{equation} 
with 
\begin{equation}
\begin{aligned}
P_{\rm m}(m_1,m_2 |{\bf \Lambda}_{\rm A}) &= C({\bf \Lambda}_{\rm A})\mathcal{PS}(m_1|\alpha_{\rm A},\delta_{\rm A},m_{\rm min,A},m_{\rm max,A}; f(m|\{x_i\},\{f_{i}\}_{\rm A}))\\
 &\times\mathcal{PS}(m_2|\alpha_{\rm A},\delta_{\rm A},m_{\rm min,A},m_{\rm max,A}; f(m|\{x_i\},\{f_{i}\}_{\rm A})) ~(m_2/m_1)^{\beta_{\rm A}}~\Theta(m_1-m_2).
\end{aligned}
\end{equation}
where $C({\bf \Lambda}_{\rm A})$ is the normalization factor, $\Theta(m_1-m_2)$ denotes the Heaviside step function ensuring $m_1>m_2$, and $\mathcal{PS}$ is the one-dimensional \textsc{PowerLaw Spline} model \citep{2022ApJ...924..101E}. We use 10 knots $\{(x_i,f_{i})\}_{i=1}^{10}$ to interpolate the perturbation function $f(m)$ of the mass distribution for first-generation BHs, and fix the locations of knots, as $\{\ln x_i\}_{i=1}^{10}$ linear in ($\ln6$, $\ln60$).

If there is single population in the spin-magnitude distribution, i.e., in the absence of the high-spin subpopulation, we can similarly model the mass and spin distributions of the {Isotropic} (2nd) subpopulation as 
\begin{equation}
\pi_{\rm I}({\bf \theta})=P_{\rm m}(m_1,m_2 |{\bf \Lambda}_{\rm I})\times \mathcal{U}(\cos\theta_1,\cos\theta_2|-1,1)\times\mathcal{G}(a_1,a_2|\mu_{\rm a,1},\sigma_{\rm a,1},0,1),
\end{equation}
and then the mass and spin distribution of the first-generation BBHs can be expressed as
\begin{equation}\label{eq:twopop}
\begin{aligned}
\pi({\bf \theta} | {\bf \Lambda}) = [P_{\rm m}(m_1,m_2 |{\bf \Lambda}_{\rm A})\times & \mathcal{G}(\cos\theta_1,\cos\theta_2|1,\sigma_{\rm t},-1,1)\times (1-r_{\rm I}) \\
+ P_{\rm m}(m_1,m_2|{\bf \Lambda}_{\rm I})\times & \mathcal{U}(\cos\theta_1,\cos\theta_2|-1,1)\times r_{\rm I}]\times\mathcal{G}(a_1,a_2|\mu_{\rm a,1},\sigma_{\rm a,1},0,1),
\end{aligned}
\end{equation}
here `A' and `I' denote the aligned and isotropic subpopulations. This model can be considered as the extend/modified version of the \textsc{Default}
spin model in \cite{2023PhRvX..13a1048A} \citep[firstly introduced by][]{2017PhRvD..96b3012T}, for this work, we allow the aligned and the isotropic subpopulations to have respective mass functions. Here after Eq.~\ref{eq:twopop} is named \textsc{Extend Default}.
{Note that the spin magnitudes in the Default spin model of \cite{2023PhRvX..13a1048A} are described by a Beta distribution. In this work, we use the truncated Gaussian for simplicity.}

However, in the presence of a high-spin subpopulation that is associated with the hierarchical mergers \citep{2024PhRvL.133e1401L}, when modeling the distribution of the 2nd part, we should consider the high-spin BHs as a distinct subpopulation, {accounting for a fraction of} $r_{\rm HS}$. In \cite{2024PhRvL.133e1401L}, we also found that there may be a fraction of near-aligned assembly in the high-spin subpopulation. Such assembly can be interpreted as the hierarchical mergers in the gas-rich environments, like disks of active galactic nucleus (AGN) \citep{2015PhRvL.115n1102G,2019PhRvL.123r1101Y}. See Section~\uppercase\expandafter{\romannumeral 9} in Supplemental Material of \cite{2024PhRvL.133e1401L} for more details. We use a half-Gaussian distribution to model the spin-orientation distribution of this assembly, hereafter labeled as the Second Aligned assembly (2A).
Therefore, the spin-orientation distribution of the high-spin subpopulation is expressed as $\mathcal{GU}(\cos\theta | \sigma_{\rm t, 2A}, \zeta_{\rm HS,2A})= (1-\zeta_{\rm HS,2A})\times \mathcal{U}(\cos\theta |-1,1)+\zeta_{\rm HS,2A} \times \mathcal{G} (\cos\theta | 1, \sigma_{\rm t,  2A}, -1, 1)$, where $\zeta_{\rm HS,2A}$ is the fraction of the aligned BHs (in the HS subpopulation), and $\sigma_{\rm t, 2A}$ is the width of $\cos\theta$ distribution of 2A assembly. Meanwhile, we use $\mathcal{GU}(\cos\theta | \sigma_{\rm t, 2A}, \zeta_{\rm LS,2A})$ to model the spin-orientation distribution of the low-spin (LS) BHs, which may be paired to the high-spin BHs \citep{2024PhRvL.133e1401L}. 
We assume the proportion between the LS-2A BHs and HS-2A BHs to be $k$, so there is a relation $k~r_{\rm HS}~\zeta_{\rm HS,2A}=(1-r_{\rm HS})~\zeta_{\rm LS,2A}$, i.e., $\zeta_{\rm LS,2A}=k~r_{\rm HS}\zeta_{\rm HS,2A}/(1-r_{\rm HS})$, with constraint of $\zeta_{\rm LS,2A}<1$. For simplicity, we set $k=3$, which is potentially associated with hierarchical mergers in AGN disks \footnote{\cite{2019PhRvL.123r1101Y} find the hierarchical mergers take a fraction of 50\%, and most higher-generation BHs are paired to the first-generation BHs. Therefore, for all the mergers in the AGN disks, the proportion between the first-generation BHs and higher-generation BHs is $\sim3$.}
Therefore the distribution of the BHs in the 2nd part is expressed as

\begin{equation}
\begin{aligned}
\pi_{\rm 2nd,BH}(m,a, \cos\theta &| {\bf \Lambda}_{\rm 2nd,BH}) = \mathcal{PS}_{\rm LS}(m|\alpha_{\rm 2nd,LS},\delta_{\rm 2nd,LS}, m_{\rm min,2nd,LS},m_{\rm max,2nd,LS}; f_{\rm 2nd, LS}(m|\{x_i\},\{f^i_{\rm 2nd,LS}\}))\\
 \times &\mathcal{G}(a|\mu_{\rm a,1},\sigma_{\rm a,1},0,1) \times \mathcal{GU}(\cos\theta | \sigma_{\rm t, 2A},\zeta_{\rm LS,2A}) \times (1-r_{\rm HS})\\
 + &\mathcal{PS}_{\rm HS}(m|\alpha_{\rm 2nd,HS},\delta_{\rm }, m_{\rm min,2nd,HS},m_{\rm max,2nd,HS}; f_{\rm 2nd,HS}(m|\{x_i\},\{f^i_{\rm 2nd,HS}\}))\\
  \times &\mathcal{G}(a|\mu_{\rm a,2},\sigma_{\rm a,2},0,1) \times \mathcal{GU}(\cos\theta | \sigma_{\rm t, 2A}, \zeta_{\rm HS,2A})\times r_{\rm HS}.
\end{aligned}
\end{equation}

Here we use another \textsc{PowerLawSpline} to model the mass function of higher-generation BHs, and for simplicity, we use 7 knots $\{(x_i,f_{i,{\rm 2nd,HS}})\}_{i=1}^{7}$ to interpolate the perturbation function $f_{\rm  2nd,HS}(m)$, and fix the locations of knots $\{x_i\}_{i=1}^{7}$ to be linear in the log space of (20, 80) $M_{\odot}$, as indicated by \cite{2024PhRvL.133e1401L}.

Then the mass and spin distribution of the 2nd subpopulation can be expressed as
\begin{equation}
\begin{aligned}
\pi_{\rm 2nd}&(m_1,m_2 ,a_1, a_2, \cos\theta_1,\cos\theta_2 | {\bf \Lambda}_{\rm 2nd})=\\
 B({\bf \Lambda}_{\rm 2nd})\times & \pi_{\rm 2nd,BH}(m_1,a_1, \cos\theta_1 | {\bf \Lambda}_{\rm 2nd,BH}) \times \pi_{\rm 2nd,BH}(m_2,a_2, \cos\theta_2 | {\bf \Lambda}_{\rm 2nd,BH})\times (m_2/m_1)^{\beta_{\rm 2nd}}~\Theta(m_1-m_2).
\end{aligned}
\end{equation}
where $B({\bf \Lambda}_{\rm 2nd})$ is the normalization factor. Therefore, in the presence of the hierarchical mergers, the total population is 
\begin{equation}\label{eq:threepop}
\pi(\boldsymbol{\theta} | {\bf \Lambda}) = \pi_{\rm 1st}(\boldsymbol{\theta} | {\bf \Lambda}_{\rm 1st})\times(1-r_{\rm 2nd})+ \pi_{\rm 2nd}(\boldsymbol{\theta} | {\bf \Lambda}_{\rm 2nd})\times r_{\rm 2nd},
\end{equation}
 where $\pi_{\rm 1st}(\boldsymbol{\theta} | {\bf \Lambda}_{\rm 1st})$ is the distribution of the field BBHs as expressed as Eq.~\ref{eq:onepop}. Hereafter, the Eq.~\ref{eq:threepop} is named \textsc{Main Model} in our work.
 {We summarize the analytical functions and their parameters of the \textsc{Main Model} in Tab.~\ref{tab:models}, and list the descriptions and priors for the parameters in Tab.~\ref{app:prior}.}
 Note that it is important to introduce another subpopulation for the high-spin events, otherwise the results will be changed, see Appendix~\ref{two_full}.
 We assume that the merger rate density increases with redshift, i.e., $\mathcal{R}(z)\propto(1 + z)^{\gamma}$, and for simplicity, in all the models we fix $\gamma=2.7$ as reported in \cite{2023PhRvX..13a1048A}.

\begin{table*}[htpb]
\begin{ruledtabular}
\caption{Summary of the functions used in the  \textsc{Main Model} and their parameters}\label{tab:models}
\begin{tabular}{c|c|c|c}

\multirow{2}{*}{Sub-populations}  & \multirow{2}{*}{1st} & \multicolumn{2}{c}{2nd}  \\
\cline{3-4}
&&  low spin & high spin \\
\cline{1-4}
\multirow{3}{*}{component mass} & \multirow{2}{*}{$\mathcal{PS}(m|\alpha_{\rm 1st},\delta_{\rm 1st},m_{\rm min,1st},$}   & \multirow{2}{*}{$\mathcal{PS}(m|\alpha_{\rm 2nd,LS},\delta_{\rm 2nd,LS},m_{\rm min,2nd,LS},$} &\multirow{2}{*}{$\mathcal{PS}(m|\alpha_{\rm 2nd,HS},\delta_{\rm 2nd,HS},m_{\rm min,2nd,HS},$} \\
& \multirow{2}{*}{$m_{\rm max,1st}; \{f_{\rm 1st}^j\}_{j=2}^{9})$} & \multirow{2}{*}{$m_{\rm max,2nd,LS}; \{f_{\rm 2nd,LS}^j\}_{j=2}^{9})$} & \multirow{2}{*}{$m_{\rm max,2nd,HS}; \{f_{\rm 2nd,HS}^j\}_{j=2}^{7})$} \\
&&&\\
\cline{1-4}

\multirow{2}{*}{spin magnitude} &\multirow{2}{*}{$\mathcal{G}(a|\mu_{\rm a,1},\sigma_{\rm a,1},0,1)$}&\multirow{2}{*}{$\mathcal{G}(a|\mu_{\rm a,1},\sigma_{\rm a,1},0,1)$}  & \multirow{2}{*}{$\mathcal{G}(a|\mu_{\rm a,2},\sigma_{\rm a,2},0,1)$} \\
&&&\\
\cline{1-4}

\multirow{3}{*}{cosine spin tilted angle} & \multirow{3}{*}{$\mathcal{G}(\cos\theta|1,\sigma_{{\rm t,1st}}, -1,1)$} &  \multirow{2}{*}{$\mathcal{GU}(\cos\theta | \sigma_{\rm t, 2A},\zeta_{\rm LS,2A})$, with} &  \multirow{3}{*}{$\mathcal{GU}(\cos\theta | \sigma_{\rm t, 2A}, \zeta_{\rm HS,2A})$}\\
&&\multirow{2}{*}{$\zeta_{\rm LS,2A} \equiv 3~r_{\rm HS}\zeta_{\rm HS,2A}/(1-r_{\rm HS})<1$}&\\
&&&\\
\cline{1-4}

\multirow{2}{*}{mixture fractions} &\multirow{2}{*}{$1-r_{\rm 2nd}$}&$1-r_{\rm HS}$&$r_{\rm HS}$\\
\cline{3-4}
&& \multicolumn{2}{c}{$r_{\rm 2nd}$} \\
\cline{1-4}

\multirow{2}{*}{pairing function} & \multirow{2}{*}{$(m_2/m_1)^{\beta_{\rm 1st}}~\Theta(m_1-m_2)$} & \multicolumn{2}{c}{\multirow{2}{*}{$(m_2/m_1)^{\beta_{\rm 2nd}}~\Theta(m_1-m_2)$}}  \\
&\\

\end{tabular}
\tablenotetext{}{{\bf Note.} $\mathcal{GU}$ is a combination of the aligned ($\mathcal{G}$) and the isotropic  ($\mathcal{U}$) assemblies, as defined in the text.}
\end{ruledtabular}
\end{table*}

\begin{table*}[htpb]
\begin{ruledtabular}
\caption{hyperparameter and Priors for the analysis with full BBH catalog}\label{app:prior}
\begin{tabular}{ccccc}
\multirow{2}{*}{descriptions}   & \multirow{2}{*}{parameters}  & \multicolumn{3}{c}{priors}  \\
\cline{3-5}
&&1st & \multicolumn{2}{c}{2nd} \\
&& & LS & HS \\
\cline{1-5}
power-law slopes of the component-mass functions & $\alpha_{\rm 1st} / \alpha_{\rm 2nd,LS}/\alpha_{\rm 2nd,HS}$ & U(-8,8) & U(-8,8) & U(-8,8)\\
smooth scales of the lower-mass edge& $\delta_{\rm 1st} / \delta_{\rm 2nd,LS}/ \delta_{\rm 2nd,HS} [M_{\odot}]$ &U(0,10)&U(0,10) & 0\\
 minimum mass cut off &$ m_{{\rm min,1st}} / m_{{\rm min,2nd,LS}} / m_{{\rm min,2nd,HS}} [M_{\odot}]$ & U(2,10)&U(2,10)&U(20,50)\\
 maximum mass cut off &$ m_{{\rm max,1st}} / m_{{\rm max,2nd,LS}} / m_{{\rm max,2nd,HS}}[M_{\odot}]$ & U(20,100) &U(20,100)&U(60,100)\\
y-value of the spline interpolant knots &$\{f_{\rm 1st}^j\}_{j=2}^{9} / \{f_{\rm 2nd,LS}^j\}_{j=2}^{9} / \{f_{\rm 2nd,HS}^j\}_{j=2}^{7}$&$\mathcal{N}(0,1)$ &$\mathcal{N}(0,1)$ &$\mathcal{N}(0,1)$\\
 \cline{1-5}
 power-law slopes of the mass ratio pairing & $\beta_{\rm 1st} / \beta_{\rm 2nd}$ & U(-8,8) & \multicolumn{2}{c}{U(-8,8)} \\
width of the $\cos\theta_{1,2}$ distribution & $\sigma_{{\rm t,1st}} /\sigma_{{\rm t,2A}} $  & U(0.1, 4) &  \multicolumn{2}{c}{U(0.1,1)} \\
fraction of the second subpopulation (2nd-Subpop.)& $r_{\rm 2nd}$ & - &\multicolumn{2}{c}{U(0,1)}\\
fraction of the HS BHs in the 2nd-Subpop.& $r_{\rm HS}$ & - &\multicolumn{2}{c}{U(0,1)} \\
fraction of the aligned assembly in HS BHs& $\zeta_{\rm HS,2A}$ & - &\multicolumn{2}{c}{U(0,1)} \\
the proportion between the LS-2A and HS-2A BHs& $k$ & - &  \multicolumn{2}{c}{3}\\
constraint &   \multicolumn{4}{c}{$\zeta_{\rm LS,2A} \equiv k~r_{\rm HS}\zeta_{\rm HS,2A}/(1-r_{\rm HS})<1$}\\
 \cline{1-5}
central value of spin-magnitude distribution& $\mu_{{\rm a,1}}/\mu_{{\rm a,2}}$ &\multicolumn{2}{c}{ U(0,1) }& U(0,1)  \\
width of spin-magnitude distribution& $\sigma_{{\rm a,1}}/\sigma_{{\rm a,2}}$ & \multicolumn{2}{c}{ U(0.05, 0.5)  }& U(0.05, 0.5) \\
 \cline{1-5}
local merger rate density & ${\rm log}_{10}R_0[{\rm Gpc^{-3}yr^{-1}}]$ &\multicolumn{3}{c}{U(0,3)} \\
power-law slope of the merger-rate evolution & $\gamma$  &\multicolumn{3}{c}{2.7} \\
\end{tabular}
\tablenotetext{}{{\bf Note.} Here, `U' means the uniform distribution and `$\mathcal{N}(0,1)$' means the normal distribution.}
\end{ruledtabular}
\end{table*}

\section{Results}\label{sec:result}
With our \textsc{Main Model}, we again identify the a subpopulation of BHs with distinctive spin-magnitude distribution ($\sim 0.7$), consistent with the higher-generation BHs, which was firstly reported in our previous work \citep{2024PhRvL.133e1401L}, and was recently conformed by other analysis \citep[e.g.,][]{2024arXiv240601679P}. See Appendix~\ref{app:2G}, for the distribution of the first-generation and higher-generation BHs. Following, we mainly focus on the two subpopulations of BBHs in the low-spin population. 

\subsection{{Distributions of the subpopulations}}
\begin{figure}
	\centering  
\includegraphics[width=0.9\linewidth]{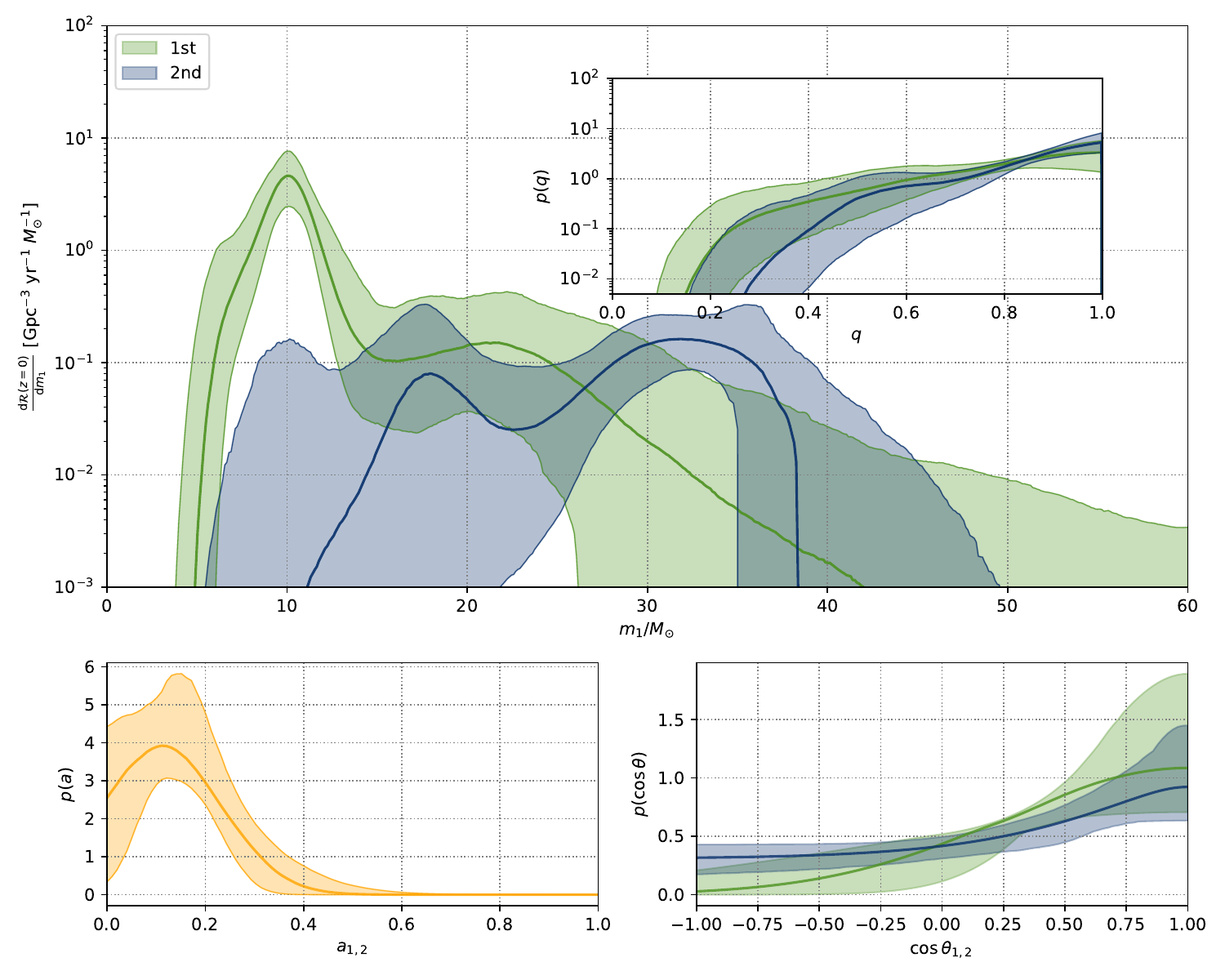}
\caption{Reconstructed primary-mass, spin-magnitude, and cosine-tilt-angle distributions of the low-spin BBHs for each subpopulation inferred with \textsc{Main Model}; the upper panel shows the differential local-universe merger rate as a function of primary mass. The solid curves are the mean values and the colored bands are the 90\% credible intervals; the insets are for the mass-ratio distributions of the two subpopulations.}
\label{fig:mass_spin}
\end{figure}

Figure \ref{fig:mass_spin} displays the distributions of the two subpopulations, {i.e., the Aligned (1st) and the Isotropic (2nd)}, in the low-spin BBHs inferred by the \textsc{Main Model}.
Note the spin-orientation distribution of the 2nd subpopulation is not fully isotropic, but there is an over density at $\cos\theta\sim1$, which is contributed by the BHs that are paired with the aligned high-spin BHs. 
The $\sim 10 M_{\odot}$ peak and the $\sim35 M_{\odot}$ peak in the primary-mass distribution, which were previously found by various approaches \citep{2021ApJ...913L...7A,2021ApJ...913L..19T,2021ApJ...917...33L,2022ApJ...924..101E}, may be dominated by the Aligned and the Isotropic subpopulations, respectively.  
Interestingly, the mass distribution of the Isotropic (2nd) subpopulation is nicely consistent with the mass distribution of BBHs in globular clusters by simulation \citep{2023MNRAS.522..466A}.
Figure~\ref{fig:three_pop_corner} displays the posterior distribution of the main parameters describing the two subpopulations; hereafter, all the values are for 90\% credible level. The {Aligned (1st)} subpopulation accounts for a fraction of $88\%^{+6\%}_{-10\%}$ of the mergers, and the width of the $\cos\theta_{1,2}$ distribution is $\sigma_{\rm t}=0.74^{+0.58}_{-0.32}$, which is smaller/tighter than the result inferred by the \textsc{Default} model \citep{2021ApJ...913L...7A,2023PhRvX..13a1048A}. 
The upper-mass cutoffs of the {1st and the 2nd-low-spin} subpopulations are $m_{\rm max,1st}=48.67^{+42.45}_{-22.68}M_{\odot}$ and $m_{\rm max,2nd,LS}=38.38^{+12.60}_{-3.35}M_{\odot}$, {which are consistent with the expectations of the (pulsational) pair-instability supernova ((P)PISN) explosions \citep{2017ApJ...836..244W,2021ApJ...912L..31W}. Although the $m_{\rm max,1st}$ is only loosely constrained (which is caused by the steep power-law slope of the mass function), the mass of the 99th percentile for the 1st subpopulation is constrained to be $m_{99\%}=28.59^{+14.40}_{-6.79}M_{\odot}$ (see Figure~\ref{fig:99per}).}
The spin-magnitude distribution peaks at $\mu_{\rm a}=0.11^{+0.06}_{-0.09}$ with a standard deviation of $\sigma_{\rm a}=0.12^{+0.07}_{-0.05}$, which is consistent with the first-generation (i.e., the low-spin) BHs inferred by the previous work \citep{2024PhRvL.133e1401L}.

\begin{figure}
	\centering  
\includegraphics[width=0.45\linewidth]{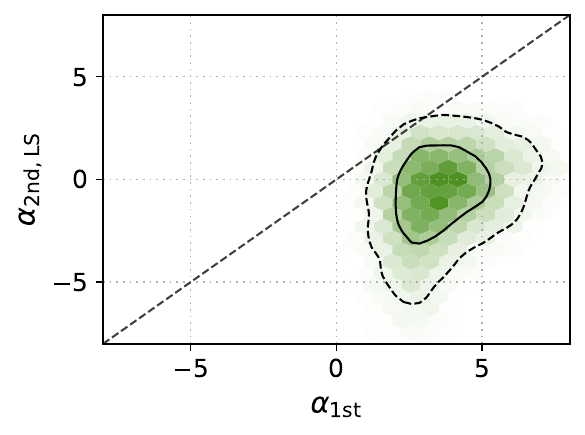}
\includegraphics[width=0.45\linewidth]{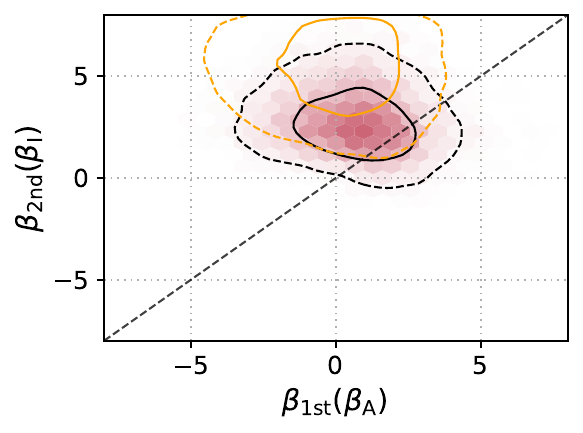}
\caption{Posterior distributions of the power-law slopes of (underlying) component-mass distributions and {pairing function} of the two subpopulations; the dashed and solid contours mark the central 50\% and 90\% posterior credible regions, respectively. The black curves are for the inference using the \textsc{Main Model} with hierarchical mergers; while the orange curves are for the inference using the \textsc{Extend Default} model without high-spin events.}
\label{fig:alpha_beta}
\end{figure}

In particular, we {find} $\alpha_{\rm 2nd,LS}<\alpha_{\rm 1st}$ at 98.5\% credible level, as shown in Figure~\ref{fig:alpha_beta} (left), i.e., the Isotropic (2nd) subpopulation has a stronger preference to produce high-mass BBH mergers than the aligned (1st) subpopulation. This result is consistent with the formation in globular clusters, particularly three-body binary formation, where the dynamical interactions will be enhanced by the total masses ($M_{\rm tot}$) of the binaries as $\propto M_{\rm tot}^{\beta_{\rm M}}$ with $\beta_{\rm M}\gtrsim4$ \citep{2016ApJ...824L..12O}. 
When inferring without the high-spin events, we find $\beta_{\rm I} > \beta_{\rm A}$ at a 95.4\% credible level, as shown in Figure~\ref{fig:alpha_beta} (the orange curves in the right panel).
{This} result is consistent with the previous investigation of the mass-ratio distribution by an independent analysis \citep{2022ApJ...933L..14L}, {which indicates} a divergence in mass-ratio distributions between low-mass and high-mass BBHs. Additionally, our findings {align with} the predictions of simulation studies \citep{2016PhRvD..93h4029R,2017MNRAS.467..524B}, which suggest that dynamical formation channels have a stronger preference for symmetric pairing. However, when inferring with high-spin events, we do not find strong evidence for $\beta_{\rm 2nd} > \beta_{\rm 1st}$. This may be because that the second subpopulation {includes} `2G+1G' systems (i.e., the hierarchical mergers with only one higher-generation BH) \citep{2024PhRvL.133e1401L}.

\subsection{Comparison with BHs in HMXBs}\label{HMXB}
Comparing the BHs in different types of sources may provide {insights} into the formation and evolution processes of compact binaries.
\citet{2022ApJ...929L..26F} compared the BHs in X-Ray binaries (XRBs) and GW sources; they found that the mass distributions of BHs in HMXBs and BBHs are consistent when accounting for the GW observational selection effects; additionally, the BHs in low-mass X-ray binaries (LMXBs) also have similar masses to BBHs with low-mass secondaries (i.e., $m_2<8M_{\odot}$). However, they found that the BHs in XRBs spin significantly faster than BBHs, leading them to suggest that the BHs in the BBHs and XRBs are like `Apple and Orange'.
On the contrary, \citet{2021arXiv211109401B} show that the BHs in XRBs and BBHs may not originate from two distinct populations, so-called `All Apples'. They suggest the observed difference in mass between the two samples arises naturally from different formation environments, {as} metallicity regulates BH mass \citep{2010ApJ...714.1217B}.
Anyway, BBHs in the GWTC-3 \citep{2019PhRvX...9c1040A,2021PhRvX..11b1053A,2021arXiv210801045T,2023PhRvX..13d1039A} may {be} a mix of dynamically formed binaries and isolated-field binaries \citep{1988ApJ...329..764L,2016A&A...588A..50M,2016MNRAS.458.2634M,2018PhRvL.121p1103F,2019PhRvL.123r1101Y,2022ApJ...933L..14L,2022ApJ...941L..39W}. The dynamically formed binaries must have undergone different evolutionary processes compared to those of the XRBs.
Therefore, it would be more effective to compare the BHs in XRBs with the BBHs from an individual channel, if the two types of formation channels are identified.

\begin{figure*}
	\centering  
\includegraphics[width=0.49\linewidth]{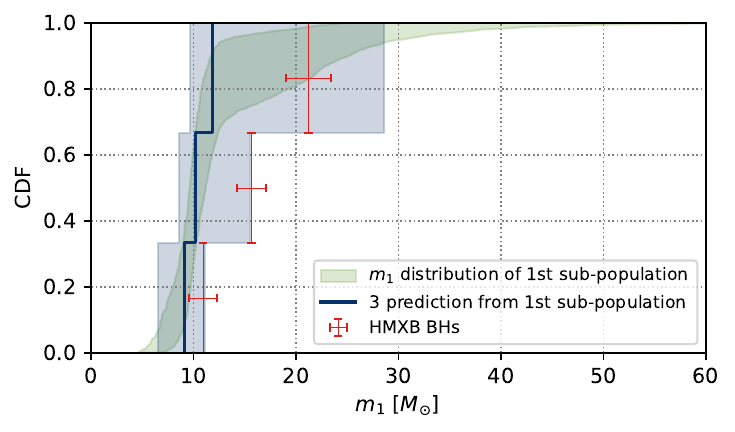}
\includegraphics[width=0.49\linewidth]{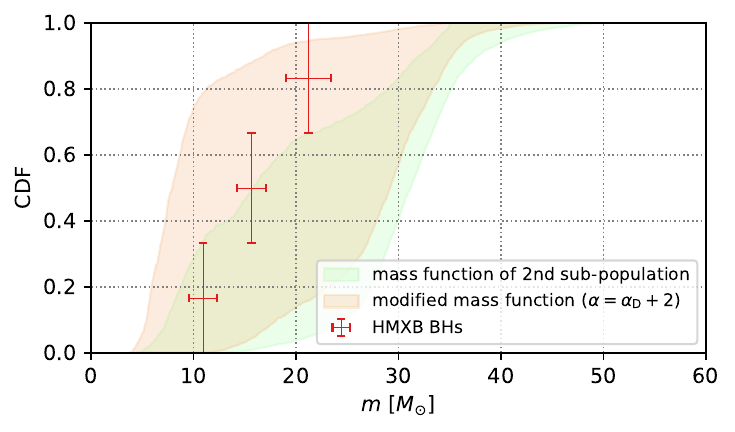}
\caption{Masses of the three observed HMXB BHs (red error bars), compared to the mass function of field BBHs (left) and dynamical BBHs (right) from GW population inference. Left: the green dashed region shows the 90\% credible interval of the inferred primary-mass function of the 1st subpopulation; the blue lines and dashed regions show the median and 90\% credible intervals of the CDF curves of three random draws from the 1st subpopulation. Right: the pale-green regions are the inferred mass function of the low-spin BHs in the 2nd subpopulation, while the sandy-brown regions are the modified mass function with $\alpha=\alpha_{\rm 2nd,LS}+2$, indicating the initial mass function of BHs in the environment for dynamical channels \citep{2016ApJ...824L..12O}.}
\label{fig:BBH_PPC}
\end{figure*}

The left panel of Figure~\ref{fig:BBH_PPC} compares the primary mass distribution of the first subpopulation, which is potentially associated with isolated-field binaries, to the mass distribution of HMXB BHs \footnote{The HMXBs adopted in this work are the same as those of \citep{2019ApJ...870L..18Q,2022ApJ...929L..26F}, including M33 X-7, Cygnus X-1, and LMC X-1, with BH masses of $15.65^{+1.45}_{-1.45}M_{\odot}$, $21.20^{+2.20}_{-2.20}M_{\odot}$, and $10.90^{+1.40}_{-1.40}M_{\odot}$, respectively.}.
The green dashed region represents the inferred primary-mass cumulative distribution function (CDF) of the field-evolution BBHs, while the blue dashed region represents the predicted CDF of three random draws from the field-evolution BBH primary mass distribution. We find that the HMXB BHs are slightly heavier than the primary BHs in the 1st subpopulation. However, the mass distributions of the HMXB BHs and the three predicted BBHs are still consistent within the Poisson uncertainty.

The spin-magnitude distribution of the BBHs inferred with our model ($\mu_{\rm a}=0.11^{+0.06}_{-0.09}$, $\sigma_{\rm a}=0.12^{+0.07}_{-0.05}$) is significantly in disagreement with the spin magnitudes of the HMXB BHs \citep{2021ARA&A..59..117R}. Such a result was also found in the comparison between the full BBH catalog and the HMXB BHs by \cite{2022ApJ...929L..26F}. 
Considering the differences in the mass functions and spin-magnitude distributions between the field-evolution BBHs and the HMXB BHs, it is possible that the HMXB BHs may have accreted significant mass, e.g., the BH mass can increase by a factor of 1.3 \citep{2022ApJ...938L..19G,2022ApJ...930...26S}; simultaneously, BHs are spun up by super-Eddington accretion under the assumption of conservative mass transfer (MT) \citep{2022RAA....22c5023Q,2022ApJ...930...26S} (or the HMXB BHs may also be rapidly spinning initially \citep{2019ApJ...870L..18Q}). Conservative MT widens the orbits of the HMXBs, thereby preventing them from merging within a Hubble time \citep{2021A&A...647A.153B,2022ApJ...933...86Z}. This process leads to wide BBHs or wide neutron star-BHs \citep{2022ApJ...938L..19G}. This scenario is also consistent with the fact that the detected HMXBs are not expected to merge within the Hubble time \citep{2011ApJ...742L...2B,2012arXiv1208.2422B,2021ApJ...908..118N,2022ApJ...938L..19G}.

We also compared the mass function of the BHs in the 2nd subpopulation, which is potentially associated with dynamical channels, to that of the HMXB BHs, as shown in Figure~\ref{fig:BBH_PPC} (right). The masses of the HMXB BHs are smaller than BHs in the dynamical channel (pale-green regions). However, when we modified the mass function of the 2nd subpopulation by adding $\alpha=\alpha_{\rm 2nd,LS}+2$, as indicated by the dynamical interactions in stellar clusters \citep{2016ApJ...824L..12O}, the results (sandy-brown regions) appeared consistent with those of the HMXB BHs.

\section{model comparison and results check}\label{sec:check}
In this section, we firstly compare the \textsc{Extend Default} to some other alternative models with some specific assumptions; and then perform mock injection studies to check the validation of the results in section~\ref{sec:result}.

In our previous work \citep{2024PhRvL.133e1401L}, other (single-population) models, such as the popular \textsc{Default} spin model (accompanied with the `PP' or `PS' mass model) \citep{2023PhRvX..13a1048A}, are disfavored compared to the multi-population models that accounts for correlation between mass and spin-magnitude distributions of BHs \citep{2024PhRvL.133e1401L}.
In this work, we have {searched for} two different {formation and evolution} channels (i.e., field binary evolution and dynamical formation), which are expected to be distinguishable by the spin-orientation distributions of BBHs \citep{2016ApJ...832L...2R,2018PhRvD..98h4036G,2017MNRAS.471.2801S,2018ApJ...854L...9F}. Since most of other models do not include the main correlation between mass and spin-magnitude distributions \citep{2024PhRvL.133e1401L}, we have to leave out the hierarchical mergers {(or the high-spin subpopulation)} from our analysis when performing model comparison. This is because in the presence of the high-spin subpopulation, the other models (e.g., the popular \textsc{Default} spin model) are disfavored compared to our main model, mostly due to their failure to fit the spin-magnitude distribution of the high-spin BHs that is found by our previous work \citep{2024PhRvL.133e1401L}. 

In practice, GW170729\_185629, GW190517\_055101, GW190519\_153544, GW190521\_030229, GW190602\_175927, GW190620\_030421, GW190701\_203306, GW190706\_222641, GW190929\_012149, GW190805\_211137, GW191109\_010717, and GW191230\_180458 are leaved out, for they all have probabilities of $>0.5$\footnote{This is a manual selection, which is not suitable for a quantitative analysis, like that carried out in Section~\ref{sec:result}. However, in this section, for qualitative analysis, this manual selection unlikely affects the conclusions. Actually, we have also analyzed with a stricter selection ($>0.3$) to ensure a purer first-generation BBH samples, i.e. leaving out more events: GW190413\_134308, GW191127\_050227, and GW200216\_220804, the conclusions in this work are unchanged.} to be hierarchical mergers (or a subpopulation containing high-spin BHs), according to \cite{2024PhRvL.133e1401L}. As is introduced in subsection~\ref{pop_model}, the \textsc{Extend Default} model (i.e., Eq.~\ref{eq:twopop}) is designed to search for subpopulations when only low-spin BBHs are present. Therefore in this section, we adopt the \textsc{Extend Default} model for model comparisons and mock data studies.
 Note that, this manipulation {is an approximation to pre-modeling (high-spin events) in the \textsc{Main Model}, and} may introduce systematic errors; but it is acceptable with currently limited data, since the inferred results of the leave-out analysis align with that inferred from the full catalog with \textsc{Main Model} (Eq.~\ref{eq:threepop}). 
We have also performed inference using the \textsc{Extend Default} with the full catalog, and found that the results are significantly changed, see Figure~\ref{fig:two_full}. What's more, the \textsc{Extend Default} model is decisively disfavored compared to the \textsc{Main Model} by a Bayes factor of $\ln\mathcal{B}=7$.

\subsection{Model comparison}
\begin{table*}[htpb]
\centering
\caption{Model comparison}\label{tab:bf}
\begin{tabular}{lcc}
\hline
\hline
Models     &  $\ln{\mathcal{B}}$    \\
\hline
\textsc{Extend Default}  & 0  \\
\textsc{Default} & -1.8 \\
Single nearly aligned & -1.2  \\
Single isotropic-spin & -5.2\\
With two independent spin-magnitude distributions & -1.8 \\
\hline
\textsc{Extend Default} ($\sigma_{\rm t}<0.5$)  & 1.5  \\
\textsc{Default} ($\sigma_{\rm t}<0.5$) & -2.3 \\
Single nearly aligned ($\sigma_{\rm t}<0.5$) & -9.8  \\
\hline
\hline
\end{tabular}
\\
\begin{tabular}{l}
Note: these log Bayes factors are relative to the \textsc{Extend Default} model in our work.
\end{tabular}
\end{table*} 

For Bayesian model comparison, $\ln\mathcal{B}>2.3$ ($\ln\mathcal{B}>3.5$) is interpreted as a strong (very strong) preference for one model over another, and $\ln\mathcal{B}>4.6$ as decisive evidence \citep{HaroldJeffreysWrited1961Theory}. We have also inferred using a model that only accounts for the single nearly aligned or isotropic-spin population (See Appendix~\ref{app:single} for details of the model).  As shown in Table~\ref{tab:bf}, the population of only isotropic-spin BBHs is disfavored, which means that merging BBHs are not solely dynamically formed in clusters. Though the single nearly aligned model can not be easily ruled out, we infer the width of $\cos\theta$ distribution $\sigma_{\rm t}> 0.7$ at 97\% ($\sigma_{\rm t}>0.6$ at 99.9\%) credible level, which is inconsistent with the isolated evolution of field binaries \citep{2016ApJ...832L...2R}.

A large $\sigma_{\rm t}$ for the truncated Gaussian distribution allows for a significant fraction of highly misaligned or anti-aligned BBHs, which poses challenges for the isolated-field channels \citep{2016ApJ...832L...2R}. Therefore, we have also inferred with a restriction of $\sigma_{\rm t}<0.5$ in our models, see Appendix~\ref{app:sigmat} for more details. As shown in Table~\ref{tab:bf}, with the restriction of $\sigma_{\rm t}<0.5$, the \textsc{Extend Default} is even more favored, while the other models are less favored.

One may also ask whether it is still possible for the nearly aligned and isotropic-spin assemblies to follow the same mass function, specifically the \textsc{Default} Spin model with a PowerLawSpline mass model. 
Note that in \citet{2023PhRvX..13a1048A} the spin-magnitude distribution is described by a Beta distribution in the \textsc{Default} model, whereas we simply use the truncated Gaussian distribution. We find this model is slightly less favored than our \textsc{Extend Default} by a Bayes factor of $\ln\mathcal{B}=-1.8$. If we restrict $\sigma_{\rm t}<0.5$, then the \textsc{Default} model is strongly disfavored by a Bayes factor of $\ln\mathcal{B}=-3.8$ comparing to the \textsc{Extend Default}. However, in this case, the mass functions of the nearly aligned and isotropic-spin assemblies become closer. Therefore, with currently available data, the scenario that nearly aligned and isotropic-spin assemblies share a single population cannot be ruled out yet.

Additionally, we have inferred, using a model that assumes the spin magnitudes of the BBHs in two subpopulations follow two independent distributions (see Appendix~\ref{app:twospin} for details), and found the spin-magnitude distributions of the two subpopulations are nearly identical as shown in Figure~\ref{fig:spin_magnitude}. Furthermore, this model is slightly disfavored compared to our \textsc{Extend Default}.  Therefore, based on the current GW data, there is no evidence that the spin magnitudes of first-generation merging BBHs differ between the nearly aligned and isotropic-spin subpopulations.

In the parameter estimation of individual GW events, the $\cos\theta_1$ is likely to be better measured than the $\cos\theta_2$ in the parameter \citep{2019PhRvX...9c1040A,2021PhRvX..11b1053A,2021arXiv210801045T,2023PhRvX..13d1039A}. Therefore, we also conduct an inference in the absence of $\cos\theta_2$, but our conclusions remain unchanged; see Appendix~\ref{app:ct1} for the results of this inference. Interestingly, the width of $\cos\theta_1$ distribution becomes smaller as $\sigma_{\rm t}=0.43^{+0.53}_{-0.26}$, which is an expected result for binaries from field evolution \citep{2016ApJ...832L...2R}.

\subsection{Mock data studies}
To test the validation of our results, we perform the injections, recovery, and hierarchical inferences of mock populations of BBH events. 
We respectively generate two kinds of mock populations: one (here after `Two pop') has population properties the same as those we have inferred from the real data, and the other (here after `One pop') does not have the feature of the spin-orientation distributions found in this work, see Appendix~\ref{app:sim} for the details. 

Then we use the \textsc{Extend Default} model to infer the underlying mock populations, respectively. We find similar features as we have found in the real data from the `Two pop' mock population, i.e., there are two subpopulations with different mass, spin-orientation, and mass-ration distributions, as shown in Figure~\ref{fig:Withcorr_sim}. We obtain the width of the $\cos\theta$ distribution of the aligned subpopulation as $\sigma_{\rm t}=0.6^{+0.23}_{-0.20}$, this value is similar to the results inferred from the real data, but larger than our injection of $\sigma_{\rm t}=0.3$. This indicates that the real $\cos\theta$ distribution for the aligned subpopulation of BBHs may be narrower than we inferred, and would be more consistent with the expectation about the isolated-field binaries \citep{2016ApJ...832L...2R}. For the `One pop' mock population, we do not find two significant subpopulations with different mass distributions, as shown in Figure~\ref{fig:Nocorr_sim}, because the spin-orientation distributions are all the same in all the mass range. These mock population studies support that the inferred features of BBH population from the real data are reliable, though we can not rule out that our analysis could be affected by other systematic biases, e.g., the biases in the waveform models used for parameter estimation. 
{Note that the mock data study serves as an approximate proxy for real data analysis, since we do not include (i.e., ideally remove) a subpopulation of high-spin events that is pre-modeled in the real data analysis with the \textsc{Main Model}.}

\subsection{Caveats}
The identification of the {Aligned (1st) and Isotropic (2nd) subpopulations} in this work, depends upon the modeling of the events in the high-mass range, which is considered as the hierarchical mergers \citep{2024PhRvL.133e1401L} for they contain BHs with significantly larger spin magnitudes ($\sim 0.7$) \citep{2017ApJ...840L..24F,2017PhRvD..95l4046G,2021NatAs...5..749G}. If we regard these high-spin events as the same population as the others (i.e., the first-generation / low-spin events), then we can not obtain the results displayed in Section~\ref{sec:result}, as shown in  Figure~\ref{fig:two_full}. This may be because the spin-magnitude distribution will influence the spin-tilt distribution \citep{2024arXiv240105613M}. Directly removing these high-spin events from the analysis makes it easier to identify the two subpopulations with different spin-orientation distributions, though this manipulation is {an approximation to pre-modeling (high-spin events) in the \textsc{Main Model}, and may introduce systematic errors}. What's more, for the \textsc{Main Model} in this work, we find it is necessary to introduce a nearly aligned assembly for the high-spin subpopulation (i.e., the hierarchical mergers, as expected by the formation channel \citep{2019PhRvL.123r1101Y,2021MNRAS.507.3362T,2024PhRvL.133e1401L} in AGN disks), otherwise the nearly aligned subpopulation will catch the events in the high-mass range, and display a wide $\cos\theta$ distribution (i.e., $\sigma_{\rm t}\sim1$), which is not expected by the traditional isolated-field evolution \citep{2016ApJ...824L..12O}. 
With the currently limited data, it is hard to identify the field and dynamical channels using the spin-orientation distribution, without the assumptions / manipulations mentioned above. Because the clusters in the mass versus spin-orientation distribution is not evident enough (as indicated by the Bayes factors, see Table~\ref{tab:bf}). Though the results presented in this work are model-dependent, we have revealed the potential subpopulations of the low-mass ($\lesssim 40M_{\odot}$) and low-spin ($\lesssim 0.4$) BBHs. A more general (with less assumption) population model will be needed in the future, when the data is significantly enriched \citep{2018LRR....21....3A}.

To identify the subpopulations with spin properties, a large per-event sample size (like 5000 posterior points per event as adopted in this work) is recommended. Because the potential subpopulations may have narrower spin-magnitude and spin-orientation distributions than those of the total population, if the per-event samples are not sufficient, then the narrow distributions will be mistakenly ruled out \citep{2023MNRAS.526.3495T}. We find the inferred spin-orientation distribution will be flatter if the per-event sample size is too small (see {Section~\uppercase\expandafter{\romannumeral 10} in Supplemental Material of \cite{2024PhRvL.133e1401L}} for more details).

We find that individual-event sampling may bias the measurement of spin-orientation distributions. As shown in the mock data studies, $\sigma_{\rm t}=0.3$ is set for the mock population; however, $\sigma_{\rm t}=0.6^{+0.23}_{-0.20}$ is recovered. \cite{2024arXiv240105613M} also suggests that the bias in the measured spin-magnitude and spin-orientation distributions is not driven by the waveform model but may be due to issues related to individual-event sampling.
These issues may lead to some measurements of $\cos\theta$-distributions not showing support for a peak at $1$ \citep[e.g.][see also Figure~\ref{fig:mact_spline} and Figure~\ref{fig:mact_spline_full}]{2023PhRvX..13a1048A,2022A&A...668L...2V,2023PhRvD.108j3009G,2023ApJ...946...16E}, which are in tension with the expectations {for} the isolated-field binaries \citep{2016ApJ...832L...2R}. 

The prior on the nearly aligned $\cos\theta$ distribution may affect the inferred mass distributions. As is shown in Figure~\ref{fig:mass_spin} and Figure~\ref{fig:Twopop_dist}, if the nearly aligned subpopulation allows a significant fraction of BBHs with $\cos\theta<0$, then it may dominate the $10M_{\odot}$ peak in the primary-mass function. However, if the nearly aligned subpopulation is restricted to $\cos\theta\gtrsim 0$, then the {isotropic subpopulation} may also significantly contribute to the $10M_{\odot}$ peak in the primary-mass function. 
Note that although the $\cos\theta$ distribution of the nearly aligned subpopulation may be not half-Gaussian (peaking at 1), there is a significant bump in [0.5, 1], as shown by Figure~\ref{fig:mut}, Figure~\ref{fig:mact_spline} and Figure~\ref{fig:mact_spline_full}. 

\section{conclusion and discussion}\label{sec:con}

With the rapidly growing catalog of GW events, the formation and evolution histories of coalescing compact binaries are being revealed through population analysis. Using data from GWTC-3 \citep{2019PhRvX...9c1040A,2021PhRvX..11b1053A,2021arXiv210801045T,2023PhRvX..13d1039A}, we explore the subpopulations of BBHs associated with two types of channels: field binary evolution and dynamical capture. When pre-modeling or excluding high-spin black holes, which are associated with hierarchical mergers \citep[see][]{2024PhRvL.133e1401L}, the remaining low-spin BBHs can be categorized into two subpopulations based on their spin orientation, mass, and mass-ratio distributions.
One subpopulation with isotropic spins, consistent with the dynamical formation channels, shows a stronger preference for symmetrical pairing functions. This subpopulation may dominate the $\sim35M_{\odot}$ peak in the primary-mass function found in the previous literature \citep{2021ApJ...913L...7A}. The other subpopulation, which has nearly aligned spins (as expected by the isolated-field evolution channels), exhibits a lesser preference for symmetrical pairing functions. This subpopulation likely dominates the $\sim10M_{\odot}$ peak in the primary mass distribution \citep{2021ApJ...913L..19T,2021ApJ...917...33L,2022ApJ...924..101E}.
However, it shows that a non-negligible fraction of the BBHs in the $\sim10M_{\odot}$ peak of the primary-mass distribution have spin-tilt angles $>90^{\circ}$. This phenomenon may be attributed to the dynamical formation channels in globular clusters or to specific pathways that produce BBHs with large misaligned spins in the isolated formation channel \citep{2021PhRvD.103f3032S}.

We have also compared the two subpopulations with the HMXB BHs. 
The primary-mass distribution of the 1st subpopulation may be lower than the mass distribution of BHs observed in HMXBs.
However, the mass distribution of the HMXB BHs is likely consistent with the 2nd subpopulation, modified by $\alpha=\alpha_{\rm 2nd,LS}+2$. Such a modification is motivated by the total-mass-dependent pairing mechanism in the dynamical formation channels \cite{2016ApJ...824L..12O}.
This indicates that the HMXB BHs may share the same mass function as the underlying BHs in the star clusters.

The study by \cite{2023ApJ...946...50B} has also analyzed the populations of BBHs with isotropic and aligned spin orientations in GWTC-3 and found no evidence that the BHs from the isotropic-spin population possess different distributions of mass ratios, spin magnitudes, or redshifts from the preferentially aligned-spin population. Although they did not investigate the mass function depending on the spin alignment as we do, they concluded that the dynamical and field channels cannot both exhibit mass-ratio distributions that strongly favor equal masses. In fact, according to \cite{2023ApJ...946...50B}, the isotropic-spin population is more likely to favor equal-mass systems (see their Figure 1), which is consistent with our results.

\cite{2021ApJ...922L...5C} firstly found that the unequal-mass BBHs have larger effective spins with GWTC-2 \citep{2019PhRvX...9c1040A,2021PhRvX..11b1053A}, which was further confirmed by \cite{2023PhRvX..13a1048A} using GWTC-3 \citep{2019PhRvX...9c1040A,2021PhRvX..11b1053A,2021arXiv210801045T,2023PhRvX..13d1039A}. Our previous work \citep{2024PhRvL.133e1401L} shows the hierarchical mergers may contribute to the anti-correlation between $q$-$\chi_{\rm eff}$ of BBHs (see {Section~\uppercase\expandafter{\romannumeral 10} in Supplemental Material of \cite{2024PhRvL.133e1401L}}). In this work, the field BBHs may also contribute to the anti-correlation between $q$-$\chi_{\rm eff}$, because they have less preference for equal-mass system and have more preference for aligned spins.

With a flexible mixture population model, \cite{2023arXiv230401288G} has also found that the $\sim10 M_\odot$ peak in the primary-mass distribution is associated with isolated binary formation. However, contrary to our results, \cite{2023arXiv230401288G} suggests that the events at $\sim35 M_\odot$ peak have spins consistent with the  $\sim10 M_\odot$ events. The difference (between our results and those of \cite{2023arXiv230401288G}) may be caused by the fact that the models in \cite{2023arXiv230401288G} have only two subpopulations when fitting the data which includes high-spin events or hierarchical mergers. One subpopulation is dominated by the high-spin events, while the other is the low-spin subpopulation, which occupy both the $10M_{\odot}$ and the $35M_{\odot}$ peaks. In contrast, this work introduces another distribution for the high-spin events or just exclude the high-spin events, making it possible to search for subpopulations among the low-spin events.
We note in the results of \cite{2023arXiv230401288G}, the \textsc{SpinPop$_{\rm A}$} of the \textsc{Isolated Peak Model} indeed has a preference for aligned spins. However, the \textsc{SpinPop$_{\rm A}$} of the \textsc{Peak+Continuum Model} shows less preference for alignment, which is caused by the contribution from $\sim35 M_\odot$ peak BHs (see Figure.~4 of \cite{2023arXiv230401288G}. This indicates that the $\sim35 M_\odot$ peak BHs may be associated with the dynamical channels.
Very recently, \cite{2024arXiv240403166R} {searched} for binary black hole subpopulations using binned Gaussian processes. The authors also {found} that the subpopulation {exhibiting}ure at $\sim30-40 M_\odot$ is associated with dynamical formation in globular clusters, which is consistent with our results.

The dynamical evolution is expected to produce more mergers with equal mass components \citep{2016PhRvD..93h4029R,2017MNRAS.467..524B,2019MNRAS.486.5008A,2021ApJ...910..152Z} because the comparable mass binaries have a higher binding energy, to form tighter systems and merge within shorter durations. However, isolated-field binaries through common envelopes may sometimes produce BBH mergers with unequal component masses, especially at lower metallicities \citep{2019MNRAS.485..889S,2022PhR...955....1M}. These predictions are likely supported by our result that the {Isotropic (2nd)} subpopulation has a stronger preference for symmetric pairing than the {Aligned (1st)} subpopulation.  
Note that some scenarios of field evolution can produce more symmetric BBHs \citep{2022PhR...955....1M}, {such as} homogeneous chemical evolution \citep{2016MNRAS.460.3545D,2016A&A...588A..50M,2022ApJ...941..179Q}, and some dynamical channels may mildly prefer unequal mass components \citep{2019ApJ...887L..36M}. {However, if these channels exist, they should not dominate the total population.}

Spin-orientation distribution of the field BBHs may also be mass-dependent, because lighter BBHs are {more easily} misaligned due to natal kicks \citep{2016ApJ...832L...2R}. We are exploring such tendency in the future with more GW data \citep{2018LRR....21....3A}. The fourth observing run (O4) of the LIGO, Virgo, and KAGRA GW detectors {is currently underway}, and the event samples are rapidly increasing at an unprecedented rate (see \url{https://gracedb.ligo.org/latest/}). More than twice as many events in O3 {are expected to be observed} in O4 \citep{2022PhR...955....1M}, and {we anticipate having} $\gtrsim200$ BBHs events in total, within one year; additionally, many other types of events may also be observed \citep{2022A&A...659A..84A,2022arXiv221201477T}. In future work, we will optimize our method to incorporate more {types} of BHs and BBH formation channels \citep{2018PhRvL.121h1306C,2022PhRvD.105h3526F,2023ApJ...955..127C,2022PhR...955....1M}, which should provide more insights and stronger constraints on subpopulation properties, with the increased GW samples.

\begin{acknowledgments}
We thank Yi-Zhong Fan for the constructive suggestions.
This work is supported by the National Natural Science Foundation of China (No. 12233011, No. 12203101 and No. 12303056), the Priority Research Program of the Chinese Academy of Sciences (No. XDB0550400), and the General Fund (No. 2023M733736, No.2024M753495) and the Postdoctoral Fellowship Program (GZB20230839) of the China Postdoctoral Science Foundation. This research has made use of data and software obtained from the Gravitational Wave Open Science Center (https://www.gw-openscience.org), a service of LIGO Laboratory, the LIGO Scientific Collaboration and the Virgo Collaboration. LIGO is funded by the U.S. National Science Foundation. Virgo is funded by the French Centre National de Recherche Scientifique (CNRS), the Italian Istituto Nazionale della Fisica Nucleare (INFN) and the Dutch Nikhef, with contributions by Polish and Hungarian institutes. The related codes of this work are published in \href{https://github.com/JackLee0214/Exploring-field-evolution-and-dynamical-capture-coalescing-binary-black-holes-in-GWTC-3/tree/main}{Exploring-field-evolution-and-dynamical-capture-coalescing-binary-black-holes-in-GWTC-3}.
\end{acknowledgments}

\vspace{5mm}

\software{Bilby \citep[version 1.1.4, ascl:1901.011, \url{https://git.ligo.org/lscsoft/bilby/}]{2019ascl.soft01011A},
          PyMultiNest \citep[version 2.11, ascl:1606.005, \url{https://github.com/JohannesBuchner/PyMultiNest}]{2016ascl.soft06005B}
          Nessai \citep[\url{https://nessai.readthedocs.io/en/latest/}]{Williams:2021qyt,Williams:2023ppp,nessai}
          }

\appendix
{
\section{The high-spin BHs}\label{app:2G}
It has been widely discussed that the hierarchical mergers contain BHs with significantly spin-magnitude and component mass distributions \citep{2017ApJ...840L..24F,2017PhRvD..95l4046G,2021NatAs...5..749G}. Evidence for hierarchical mergers has also been found using some parametric methods \citep{2021ApJ...915L..35K,2022ApJ...941L..39W}. Previously, we identified a high-spin subpopulation of BHs with a semiparametric population model \citep{2024PhRvL.133e1401L}, which is consistent with hierarchical mergers. In this work, we aim to identify the formation channels—both field and dynamical—for the low-spin (i.e., first-generation) BBHs based on their spin-orientation distributions. We find it essential to model the mass and spin distributions of the high-spin BHs; otherwise, they will influence the results of this work (see Appendix~\ref{two_full}).
With our \textsc{Main Model}, we again identify the high-spin BHs in the 2nd subpopulation, which have significantly different spin-magnitude and mass distributions from the low-spin BHs, consistent with the higher-generation BHs \citep{2021NatAs...5..749G}, see Figure~\ref{fig:2G}. The high-spin BHs constitute a fraction of $ 18^{+12}_{-9}\% $ in the 2nd subpopulation, as shown in Figure~\ref{fig:three_pop_corner}.

\begin{figure*}
	\centering  
\includegraphics[width=0.9\linewidth]{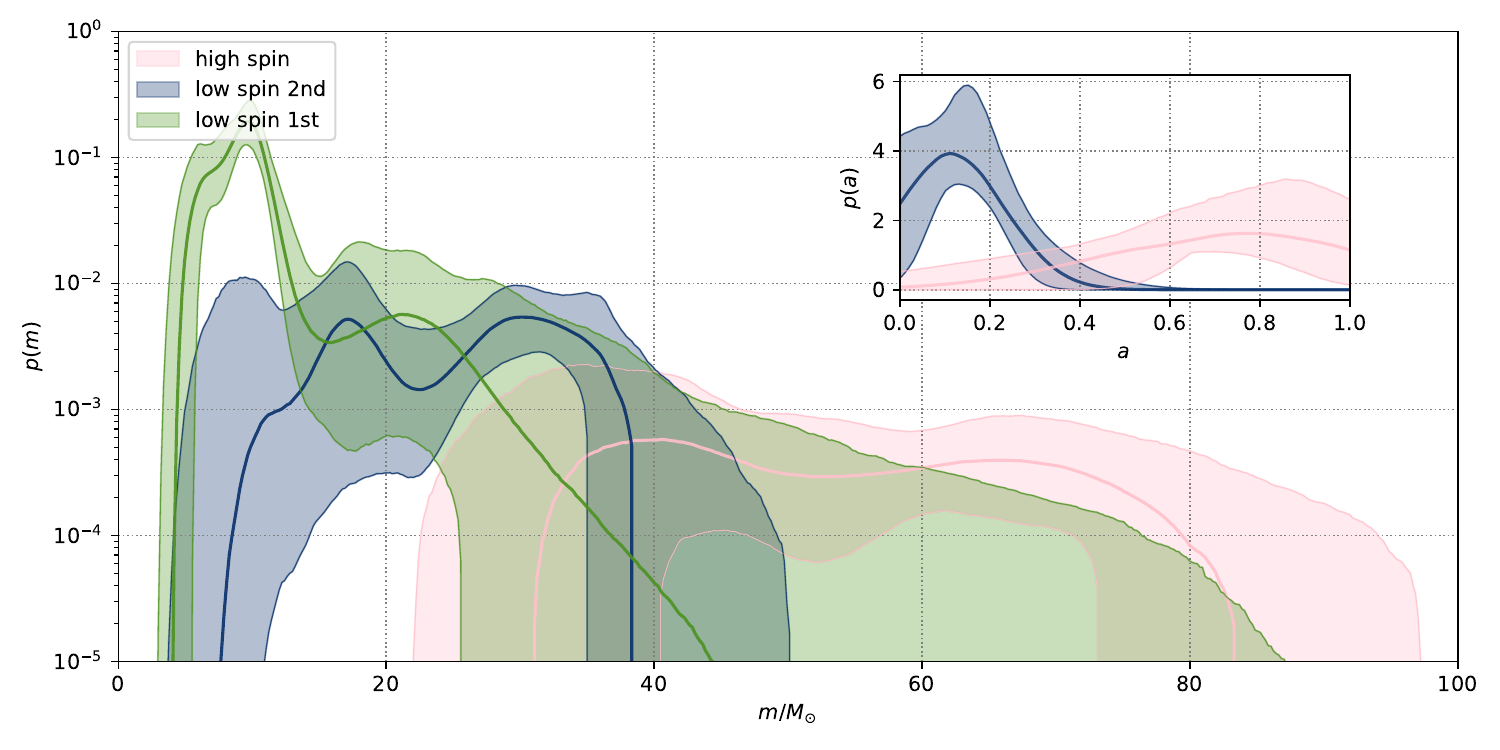}
\caption{Reconstructed {(underlying)} component-mass distributions {(before paired)}. 
The solid curves are the mean values and the colored bands are the 90\% credible intervals; the inset is for the spin-magnitude distributions. }
\label{fig:2G}
\end{figure*}
}
\begin{figure}
	\centering  
\includegraphics[width=0.98\linewidth]{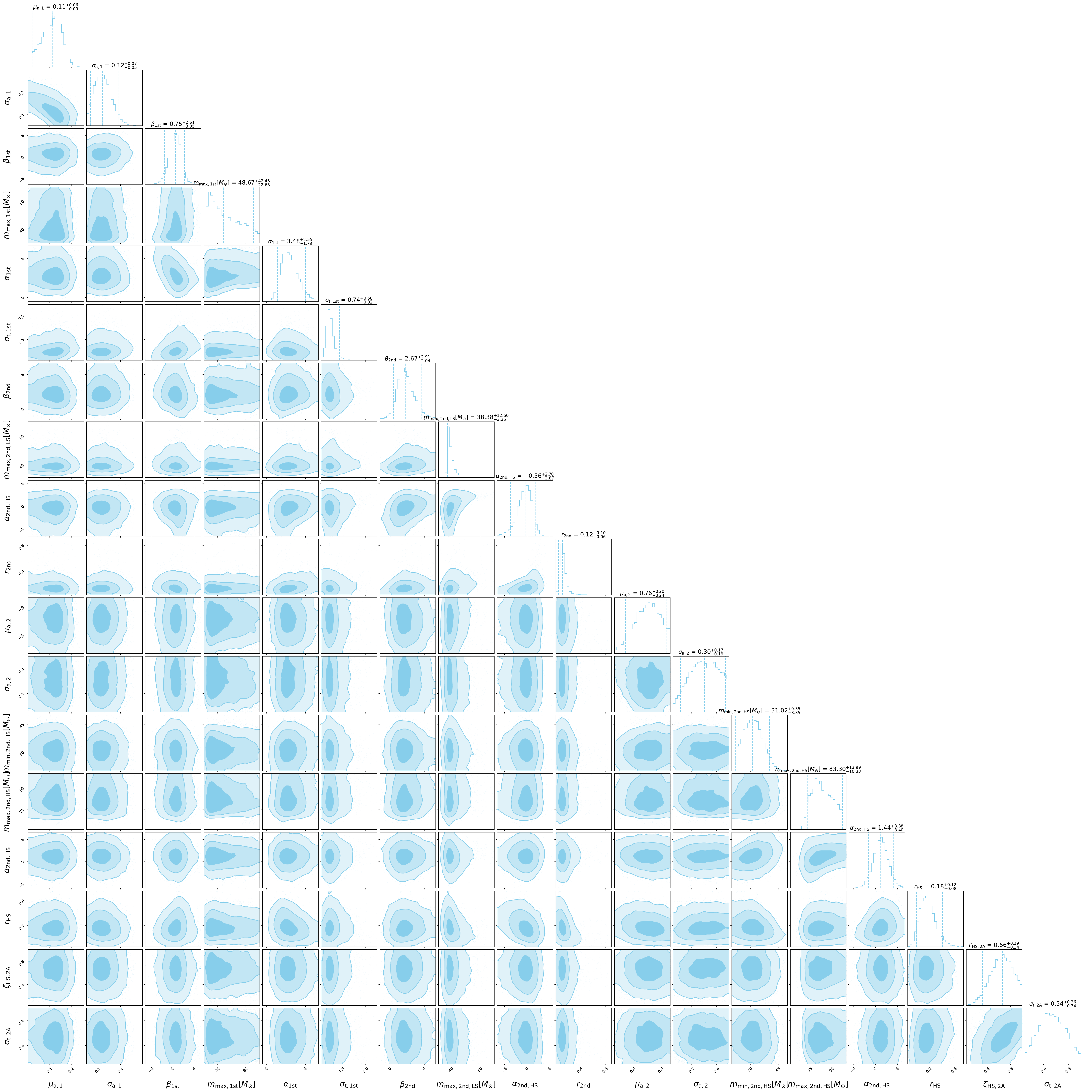}
\caption{Posterior distribution of main parameters describing the full BBH population, inferred with the \textsc{Main Model}; the dashed lines in the marginal distribution represent the 90\% credible intervals.}
\label{fig:three_pop_corner}
\end{figure}

\begin{figure*}
	\centering  
\includegraphics[width=0.4\linewidth]{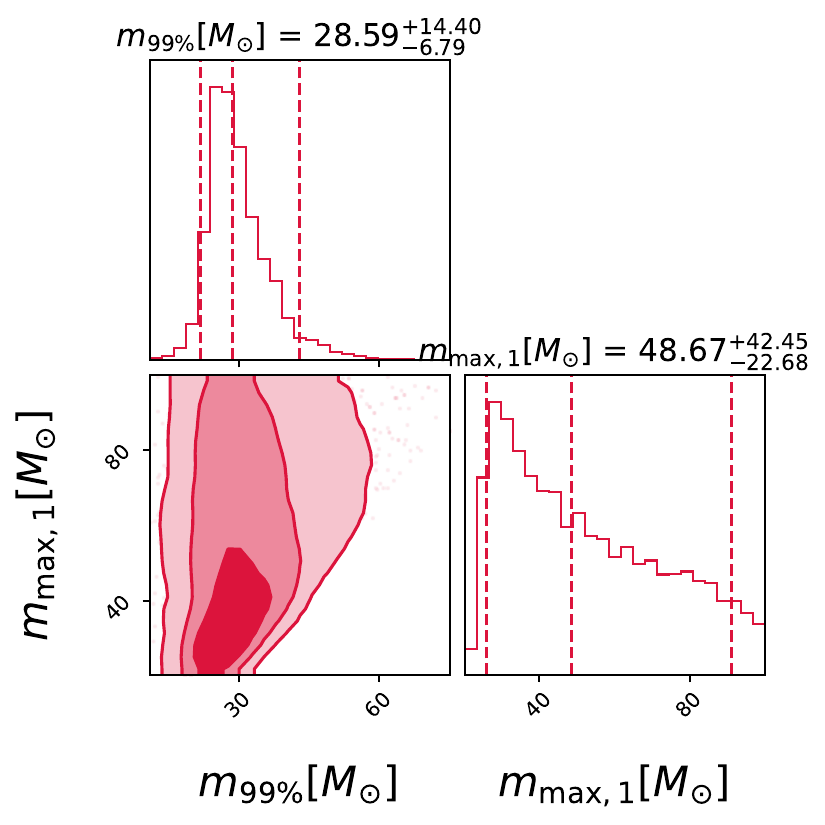}
\caption{The mass of 99th percentile and the maximum mass of the field BBHs, inferred with the \textsc{Main Model}.}
\label{fig:99per}
\end{figure*}

\section{Results of other models}\label{app:models}

\subsection{\textsc{Extend Default} model}

\begin{figure*}
	\centering  
\includegraphics[width=0.9\linewidth]{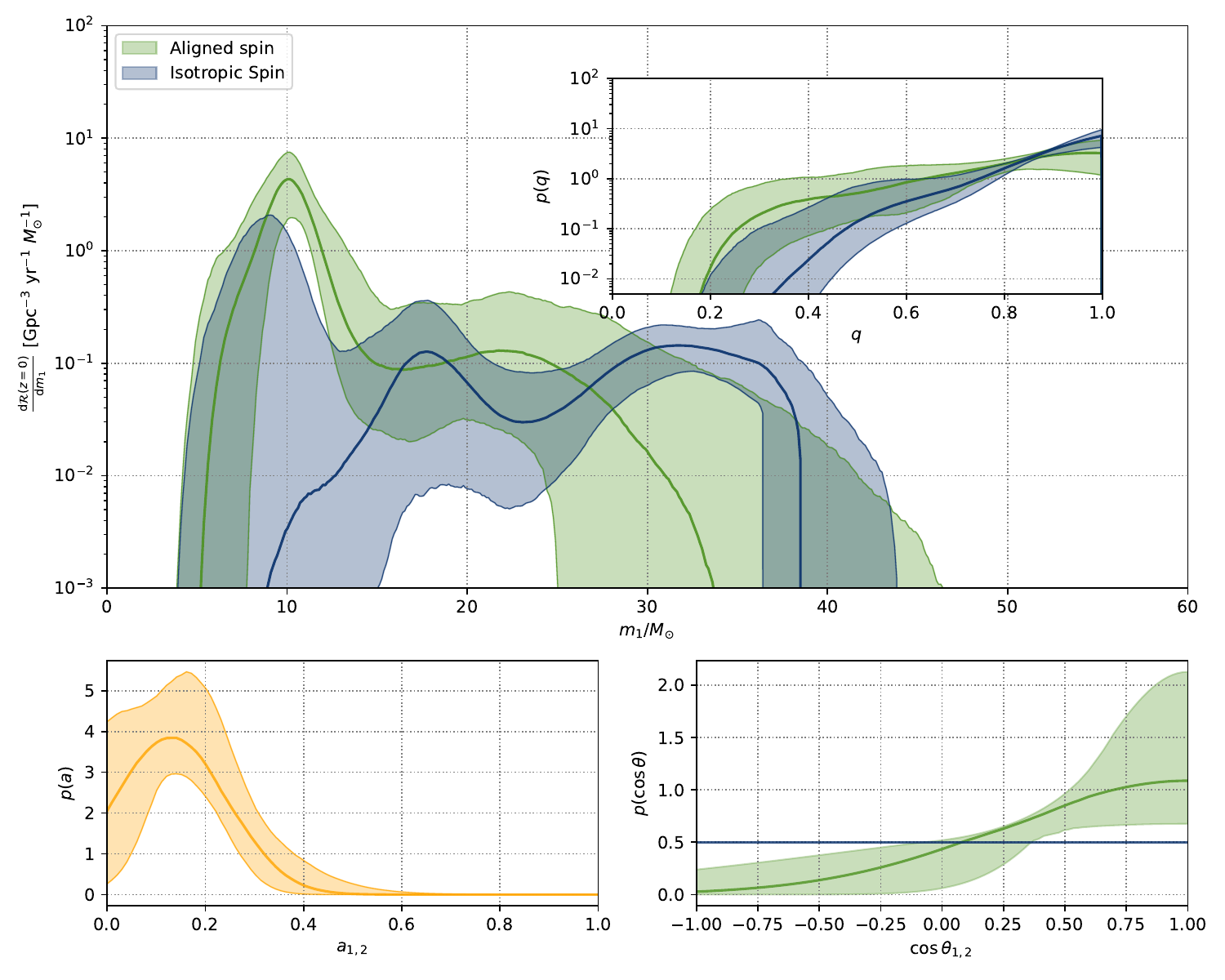}
\caption{Reconstructed primary-mass, spin-magnitude, and cosine-tilt-angle distributions of BBHs for each population, inferred with the \textsc{Extend Default} model. The upper panel shows the differential local-universe merger rate as a function of primary mass. The solid curves are the mean values and the colored bands are the 90\% credible intervals; the insets are for the mass-ratio distributions of the two subpopulations.}
\label{fig:Twopop_dist}
\end{figure*}

\begin{figure}
	\centering  
\includegraphics[width=0.9\linewidth]{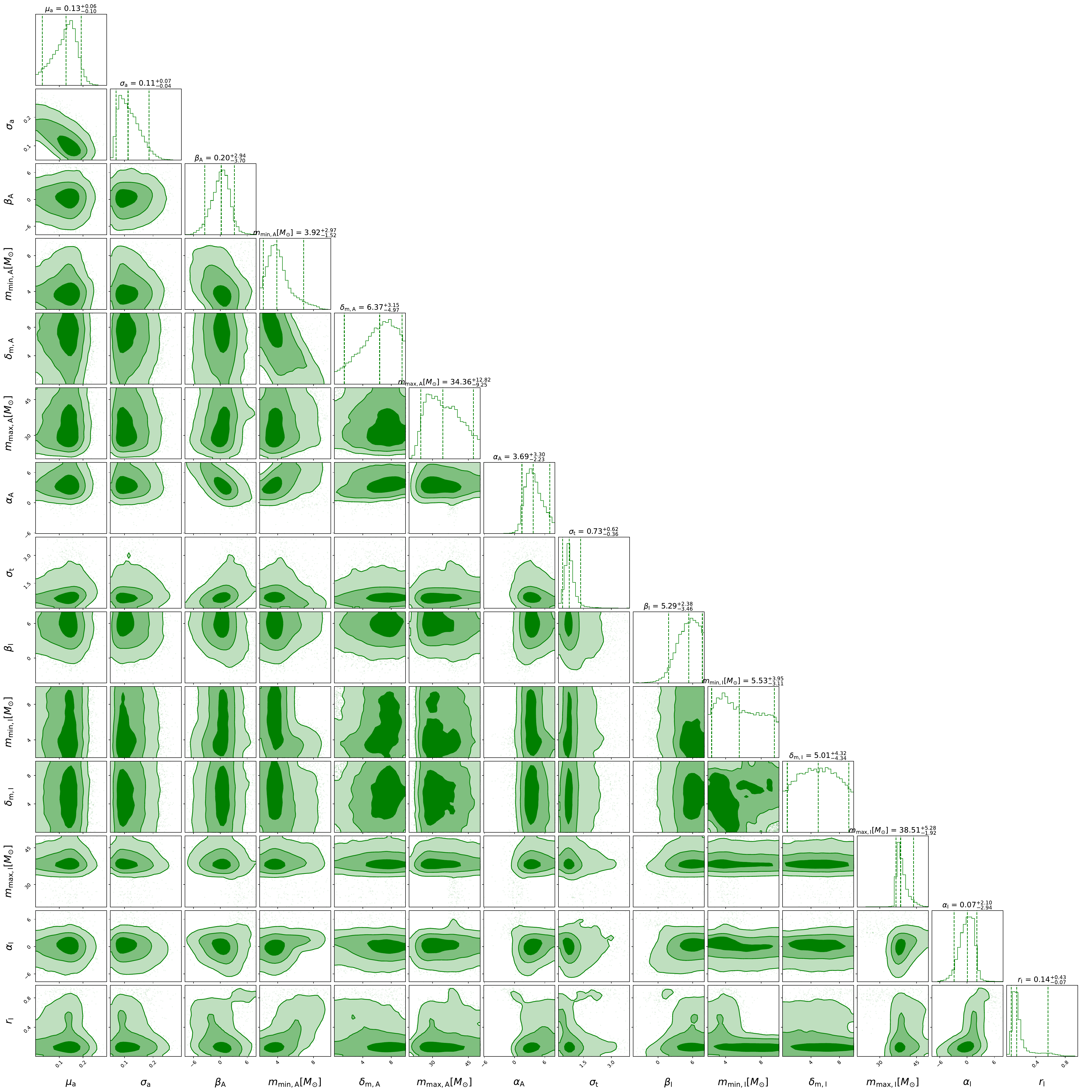}
\caption{Posterior distributions of all the hyperparameters describing the two subpopulations, inferred with the \textsc{Extend Default} model; the dashed lines in the marginal distribution represent the 90\% credible intervals.}
\label{fig:Twopop_corner}
\end{figure}

The priors for the \textsc{Extend Default} model are summarized in Table~\ref{tab:app_prior}. Figure~\ref{fig:Twopop_dist} shows the primary-mass, spin-magnitude, and cosine-tilt-angle distributions of the aligned-spin and isotropic-spin subpopulations of the first generation BBHs,  as inferred by the \textsc{Extend Default} model. And Figure~\ref{fig:Twopop_corner} presents the posterior distributions of all the hyperparameters that describe the two subpopulations.

\begin{table*}[htpb]
\begin{ruledtabular}
\caption{hyperparameter and Priors for the \textsc{Extend Default}}\label{tab:app_prior}
\begin{tabular}{cccc}
\multirow{2}{*}{descriptions}   & \multirow{2}{*}{parameters}  & \multicolumn{2}{c}{priors}  \\
\cline{3-4}
&&nearly aligned assembly & isotropic-spin assembly \\
\cline{1-4}
power-law slope of the primary mass distributions & $\alpha_{\rm A} / \alpha_{\rm I}$ & U(-4,12) & U(-4,4) \\
smooth scale of the lower-mass edge& $\delta_{\rm A} / \delta_{\rm I} [M_{\odot}]$ &U(1,10)&U(1,10)\\
 minimum mass cut off &$ m_{{\rm min,A}} / m_{{\rm min,I}} [M_{\odot}]$ & U(2,10)&U(2,10)\\
 maximum mass cut off &$ m_{{\rm max,A}} / m_{{\rm max,I}}[M_{\odot}]$ & U(20,60) &U(20,60)\\
 power-law slope of the mass ratio distribution & $\beta_{\rm A} / \beta_{\rm I}$ & U(-4,8) & U(-4,8) \\
y-value of the spline interpolate knots &$\{f_{\rm A}^j\}_{j=2}^{9} / \{f_{\rm I}^j\}_{j=2}^{9}$ &$\mathcal{N}(0,1)$ &$\mathcal{N}(0,1)$ \\
Width of the $\cos\theta_{1,2}$ distribution & $\sigma_{{\rm t}} $  & U(0.1, 4) & - \\
mixing fraction of the isotropic-spin assembly& $r_{\rm I}$ & - &{U(0,1)}\\
 \cline{1-4}
Central value of spin-magnitude distribution& $\mu_{{\rm a}}$ &\multicolumn{2}{c}{ U(0,1) } \\
Variance of spin-magnitude distribution& $\sigma_{{\rm a}}$ & \multicolumn{2}{c}{ U(0.02, 0.5)  }\\
local merger rate density & ${\rm log}_{10}R_0[{\rm Gpc^{-3}yr^{-1}}]$ &\multicolumn{2}{c}{U(0,3)}\\
power-law slope of the merger-rate evolution & $\gamma$  &\multicolumn{2}{c}{2.7} \\
\end{tabular}
\tablenotetext{}{{\bf Note.} Here, `U' means the uniform distribution and `$\mathcal{N}(0,1)$' means the normal distribution.}
\end{ruledtabular}
\end{table*}

\subsection{Inference with $\sigma_{\rm t}<0.5$}\label{app:sigmat}
 
\begin{figure}
	\centering  
\includegraphics[width=0.9\linewidth]{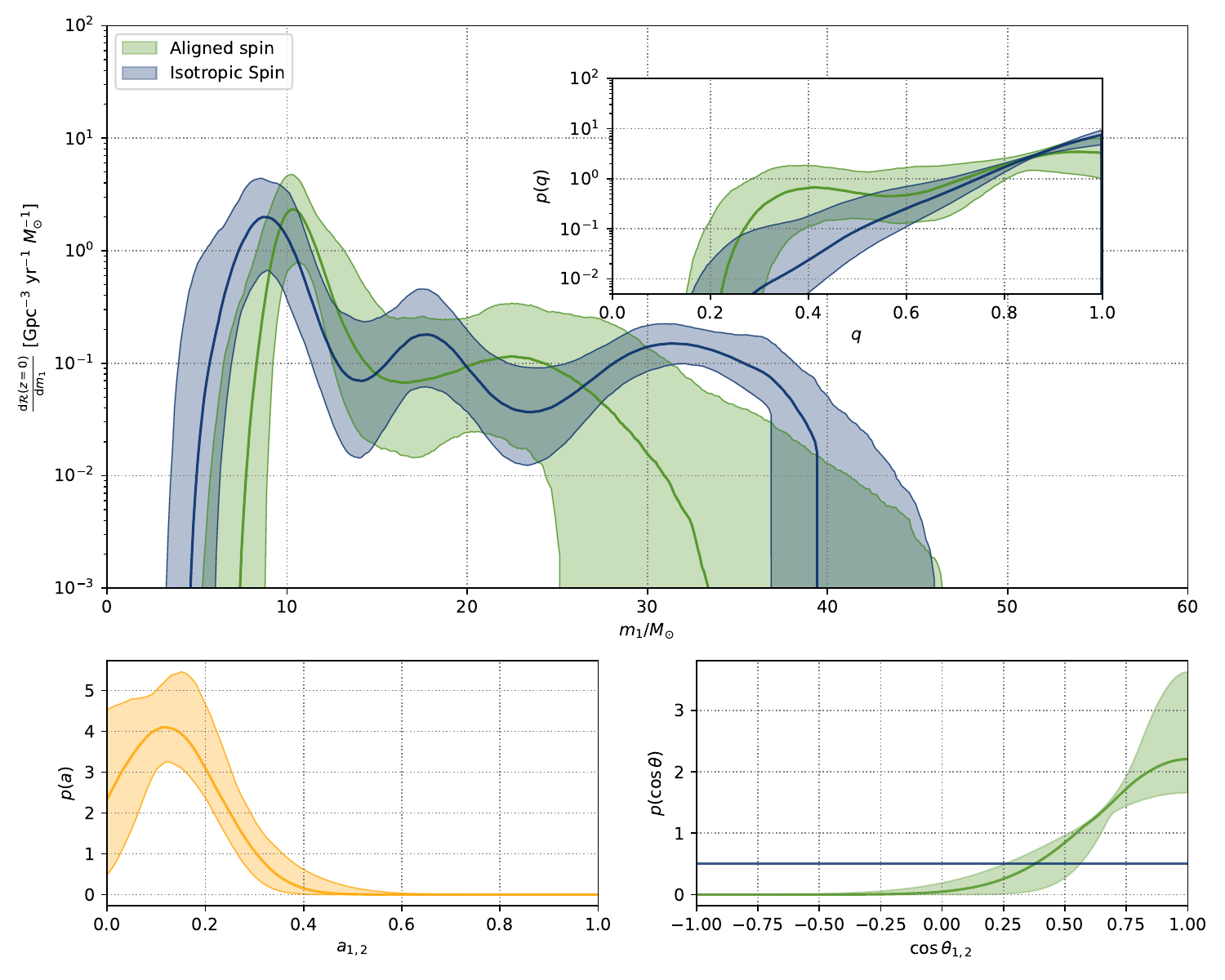}
\caption{The same as Figure~\ref{fig:Twopop_dist} but for the restriction of $\sigma_{\rm t}<0.5$.}
\label{fig:restrict}
\end{figure}

With a restriction of $\sigma_{\rm t}<0.5$, the primary-mass function of the nearly aligned subpopulation is only slightly changed, as shown in Figure~\ref{fig:restrict}. However, the primary-mass function of the isotropic-spin subpopulation is significantly changed in the low-mass range, and the fraction of the isotropic-spin subpopulation turns to $0.57^{+0.22}_{-0.23}$. What's more, the nearly aligned subpopulation has a flatter mass-ratio distribution, and the pairing function is $(m_2/m_1)^{\beta_{\rm A}}$ with $\beta_{\rm A}=-1.59^{+3.39}_{-3.62}$. The merger rate density of the isotropic-spin subpopulation is raised in the low-mass range, but it is still possible contributed by the field channel, because the lighter binaries are more likely to change their spin orientations in supernovea explosions by natal kicks than the heavier binaries \citep{2016ApJ...832L...2R}. 
Note that there is a sub-mode in the posterior distribution inferred without the restriction, as shown in Figure~\ref{fig:Twopop_corner}, which is consistent with the results inferred with $\sigma_{\rm t}<0.5$.  

\subsection{Inference with a variable $\mu_{\rm t}$}\label{app:mut}
{\cite{2022A&A...668L...2V} found no evidence for the $\cos\theta$ distribution peaking at $+1$. Here, we have also inferred with a variable $\mu_{\rm t}$ in the \textsc{Extend Default}. As show in Figure~\ref{fig:mut_corner}, the $\mu_{\rm t}$ peaks at $\gtrsim0.5$, and $\mu_{\rm t}<0$ is strongly disfavored. Additionally, the mass and spin-magnitude distributions are unchanged, and the $\cos\theta$ distribution shows a significant bump at [0.5, 1], as shown in Figure~\ref{fig:mut}.
{The difference (between our result and that of \cite{2022A&A...668L...2V}) is mainly caused by the different treatment for high-spin subpopulation, see Section~\uppercase\expandafter{\romannumeral 6} in Supplemental Material of \cite{2024PhRvL.133e1401L} for more details}.
\begin{figure*}
	\centering  
\includegraphics[width=0.9\linewidth]{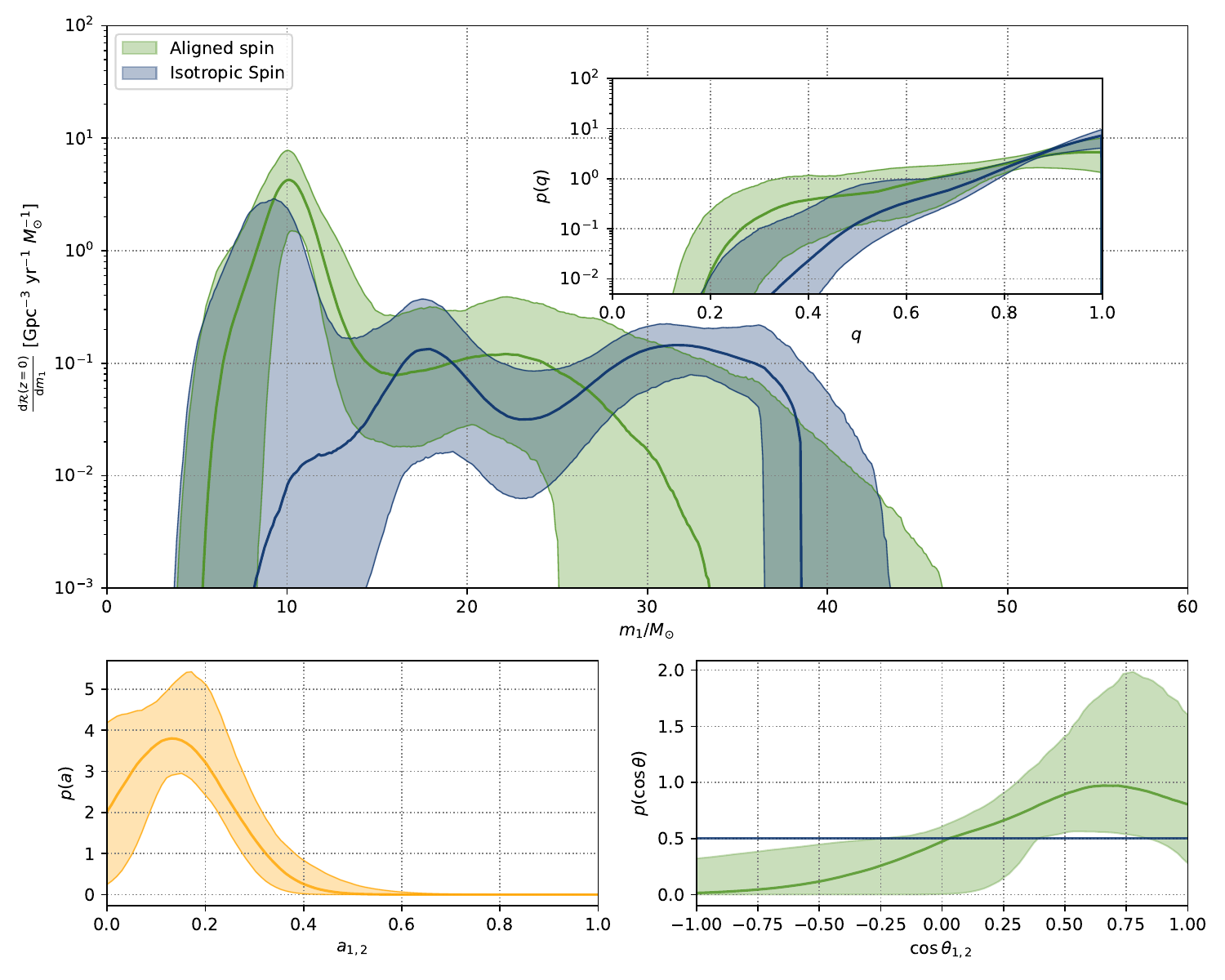}
\caption{The same as Figure~\ref{fig:Twopop_dist}, but for the case with variable $\mu_{\rm t}$.}
\label{fig:mut}
\end{figure*}

\begin{figure}
	\centering  
\includegraphics[width=0.3\linewidth]{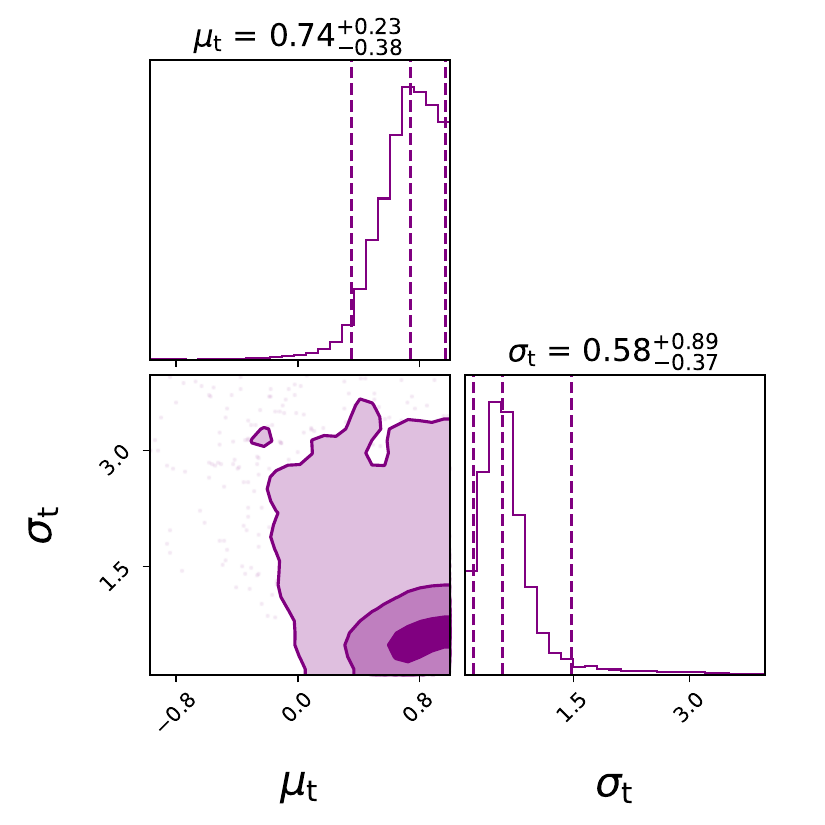}
\caption{Posterior distributions of the $\mu_{\rm t}$ and $\sigma_{\rm t}$ describing the $\cos\theta_{1,2}$ distribution of the nearly aligned subpopulation; the dashed lines in the marginal distribution represent the 90\% credible intervals.}
\label{fig:mut_corner}
\end{figure}
}

\subsection{Is single nearly aligned or isotropic-spin population also consistent with data?}\label{app:single}

\citet{2021ApJ...921L..15G} found that the inferred spin distribution is nearly aligned, which is consistent with the hypothesis that all merging binaries form via the field formation scenario.
To check out whether the a single nearly aligned or a single isotropic-spin population is still satisfied by the current GW data, we infer with the single nearly aligned model $M_{\rm A}$ and the single isotropic-spin model $M_{\rm I}$, which is expressed as 
\begin{equation}
\begin{aligned}
\pi_{\rm A}(\boldsymbol{\theta} | {\bf \Lambda}) =& \mathcal{PS}(m_1,m_2 |\alpha,\beta,\delta,m_{\rm min},m_{\rm max}; f(m_1|\{x_i\},\{f_{i}\}))\\
    &\times \mathcal{G}(\cos\theta_1,\cos\theta_2|1,\sigma_{\rm t},-1,1) \times\mathcal{G}(a_1,a_2|\mu_{\rm a},\sigma_{\rm a},0,1)\times p(z|\gamma=2.7),
\end{aligned}
\end{equation}
 and 
\begin{equation}
\begin{aligned}
\pi_{\rm I}(\boldsymbol{\theta} | {\bf \Lambda}) =& \mathcal{PS}(m_1,m_2 |\alpha,\beta,\delta,m_{\rm min},m_{\rm max}; f(m_1|\{x_i\},\{f_{i}\}))\\
    &\times \mathcal{U}(cos\theta_1,\cos\theta_2|-1,1) \times\mathcal{G}(a_1,a_2|\mu_{\rm a},\sigma_{\rm a},0,1)\times p(z|\gamma=2.7).
\end{aligned}
\end{equation}
As shown in Table~\ref{tab:bf}, the $M_{\rm I}$ is ruled out by a Bayes factor of $\ln\mathcal{B}=-5.5$, and the $M_{\rm A}$ is also slightly less favored compared to the \textsc{Extend Default}. However, if we restrict $\sigma_{\rm t}$ to be $<0.5$ as expected by the isolated-field evolution \citep{2016ApJ...832L...2R}, $M_{\rm A}$ was also rule out.

\subsection{Inference with only $\cos{\theta_1}$}\label{app:ct1}

\begin{figure}
	\centering  
\includegraphics[width=0.8\linewidth]{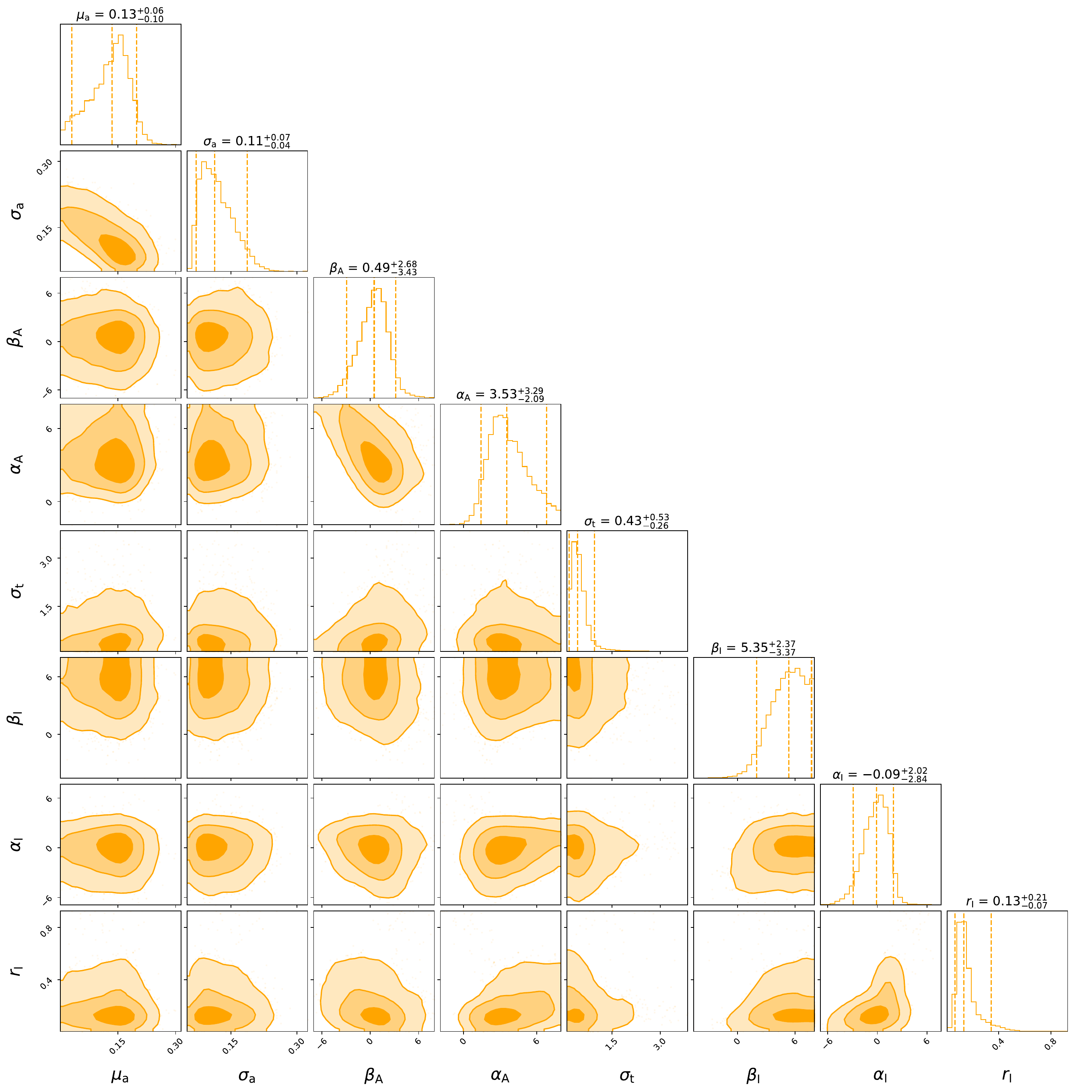}
\caption{Posterior distribution of the main hyperparameter describing the two subpopulations, inferred without $\cos\theta_2$ information.}
\label{fig:ct1}
\end{figure}

\begin{figure}
	\centering  
\includegraphics[width=0.9\linewidth]{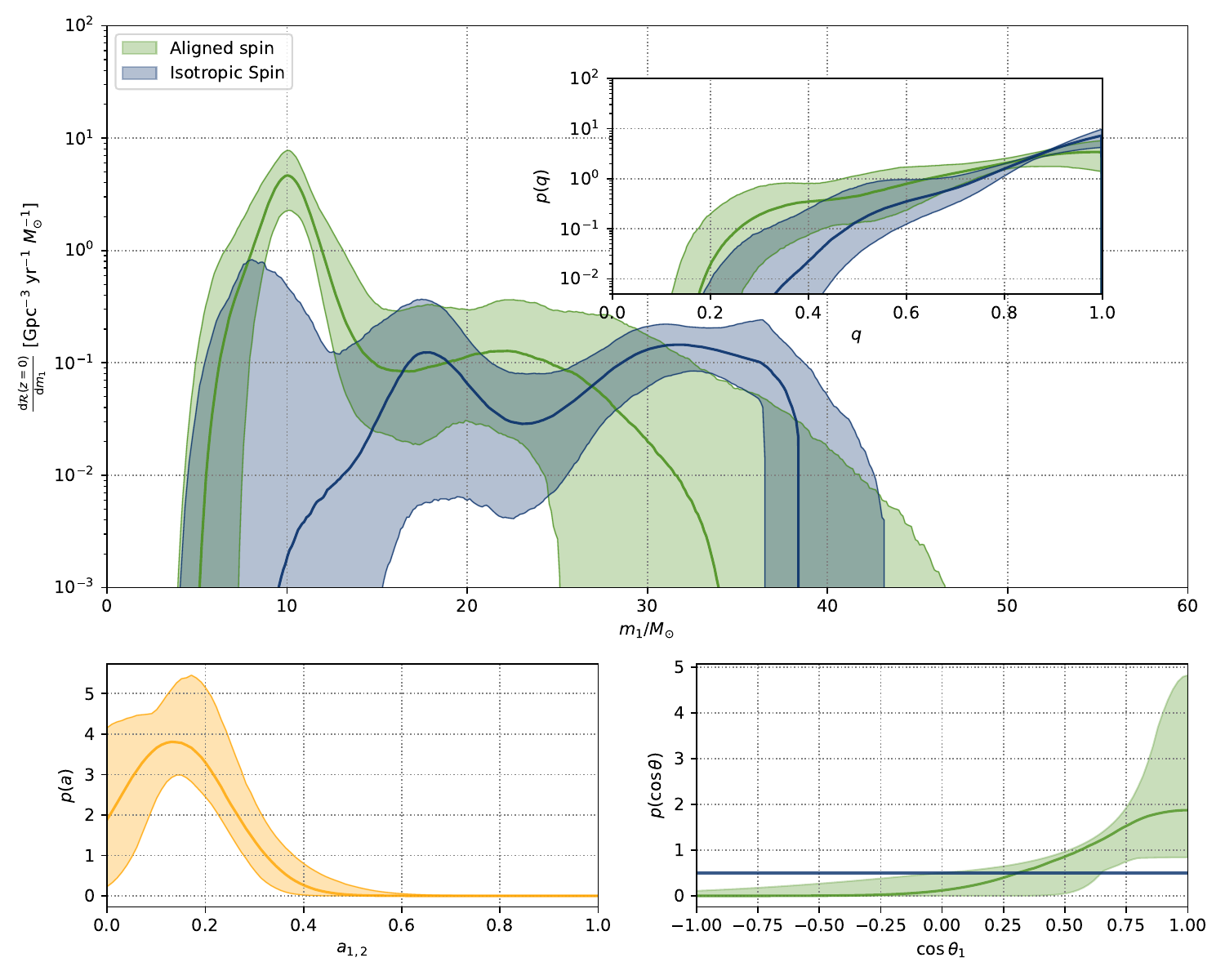}
\caption{The same as Figure~\ref{fig:Twopop_dist}, but for the inference with only $\cos\theta_1$ distribution.}
\label{fig:mact1}
\end{figure}

We also inferred with only $\cos\theta_1$ distribution, since $\cos\theta_1$ is usually constrained better than $\cos\theta_2$ in the parameter estimation for an individual event \citep{2019PhRvX...9c1040A,2021PhRvX..11b1053A,2021arXiv210801045T,2023PhRvX..13d1039A}. As displayed in Figure~\ref{fig:mact1}, the results are nearly similar to that inferred in the main text. Figure~\ref{fig:ct1} shows the posterior distribution of the main hyperparameter, we find $\sigma_{\rm t}= 0.43^{+0.53}_{-0.26}$ is smaller than that inferred in the presence of $\cos\theta_2$, while other parameters are similar to those inferred with $\cos\theta_2$.

\subsection{Is the spin-magnitude distributions of two subpopulation identical?}\label{app:twospin}
\begin{figure}
	\centering  
\includegraphics[width=0.6\linewidth]{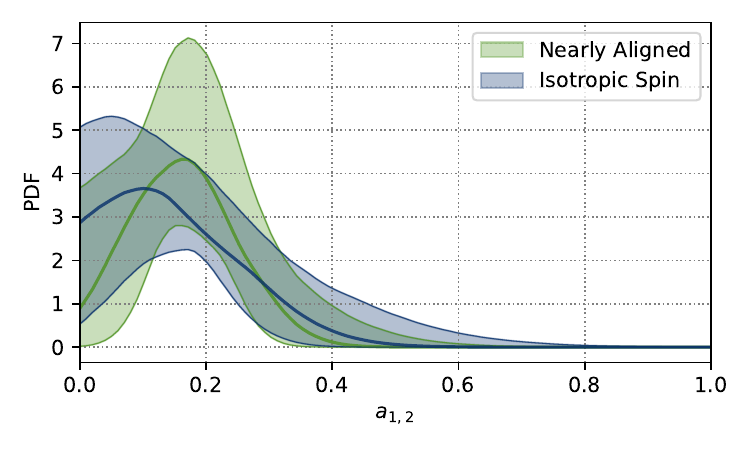}
\caption{Spin-magnitude distributions of BBHs in the nearly aligned and isotropic-spin subpopulation. The solid curves are the medians and the colored bands are the 90\% credible intervals.}
\label{fig:spin_magnitude}
\end{figure}

To check out whether the spin-magnitudes of two subpopulations have the same distribution or significantly different distributions, we have also inferred with the model with two independent truncated Gaussian distributions to describe the spin-magnitude distributions of the two subpopulations, respectively. 
\begin{equation}
\begin{aligned}
\pi(\boldsymbol{\theta} | {\bf \Lambda}) 
=  [\mathcal{PS}(m_1,m_2 &|\alpha_{\rm A},\beta_{\rm A},\delta_{\rm A},m_{\rm min,A},m_{\rm max,A}; f_{\rm A}(m_1|\{x_i\},\{f_{i,{\rm A}}\}))\\
 \times & \mathcal{G}(\cos\theta_1,\cos\theta_2|1,\sigma_{\rm t},-1,1)\times\mathcal{G}(a_1,a_2|\mu_{\rm a,A},\sigma_{\rm a,A},0,1)\times (1-r_{\rm I}) \\
+  \mathcal{PS}(m_1,m_2 &|\alpha_{\rm I},\beta_{\rm I},\delta_{\rm I},m_{\rm min,I},m_{\rm max,I};f_{\rm I}(m_1|\{x_i\},\{f_{i,{\rm I}}\}))\\
\times & \mathcal{U}(\cos\theta_1,\cos\theta_2|-1,1)\times\mathcal{G}(a_1,a_2|\mu_{\rm a,I},\sigma_{\rm a,I},0,1)\times r_{\rm I}]\\ 
\times p(z|\gamma=2.7).
\end{aligned}
\end{equation}
As is shown in Figure~\ref{fig:spin_magnitude}, the spin magnitudes from the two subpopulations are nearly identical, and the Bayes factors (see Table~\ref{tab:bf}) also show that the spin-magnitude distributions of the two subpopulations are not necessary to be different with current observation data.

\subsection{Non-parametric $\cos\theta$-distribution model}\label{app:non-para}
\begin{figure}
	\centering  
\includegraphics[width=0.9\linewidth]{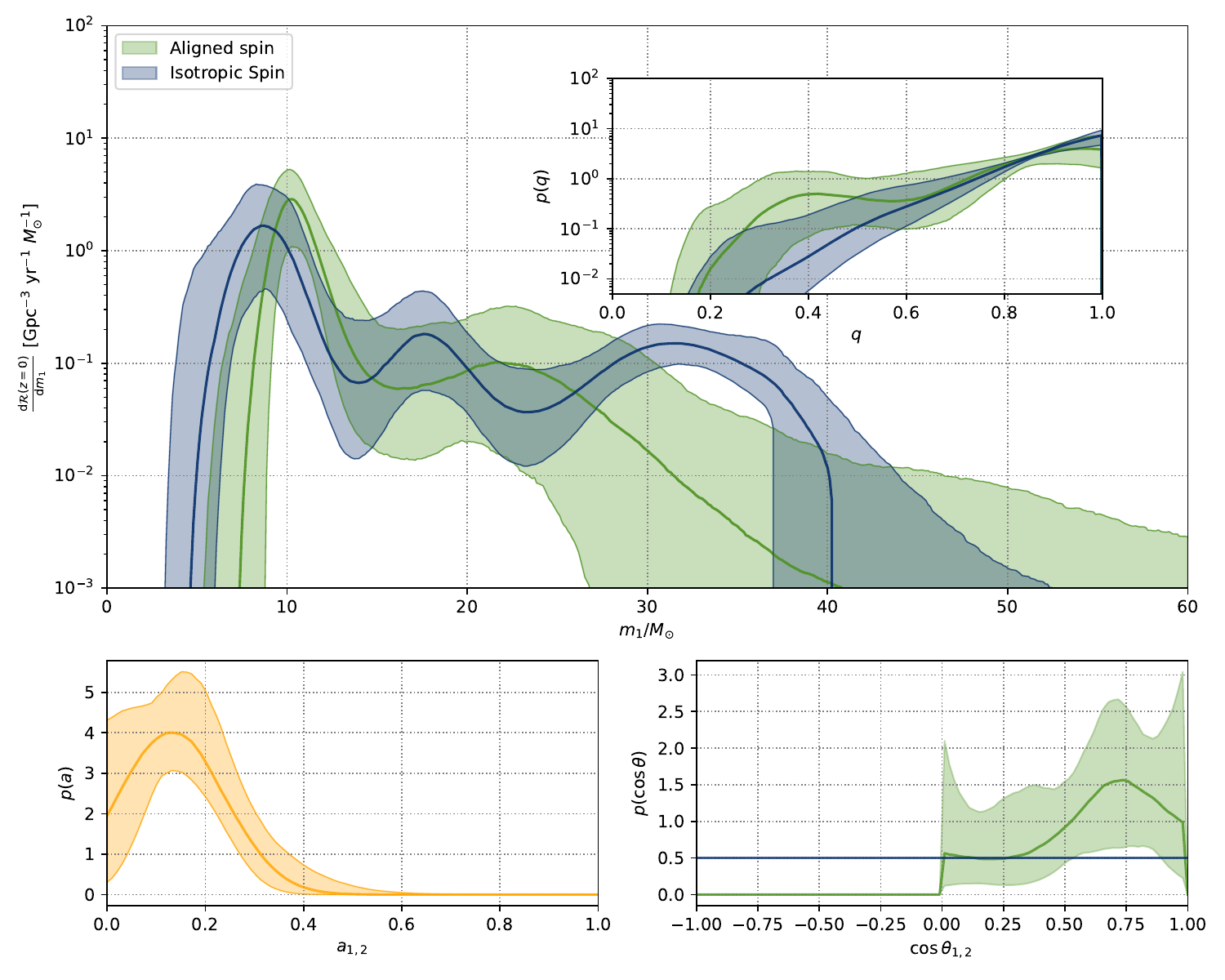}
\caption{The same as Figure~\ref{fig:Twopop_dist}, but for the inference with non-parametric $\cos\theta$-distribution in $[0,1]$. The mass distributions of the two subpopulations are rather similar to those inferred using the \textsc{Extend Default} model with $\sigma_{\rm t}<0.5$.}
\label{fig:mact_spline}
\end{figure}

\begin{figure}
	\centering  
\includegraphics[width=0.9\linewidth]{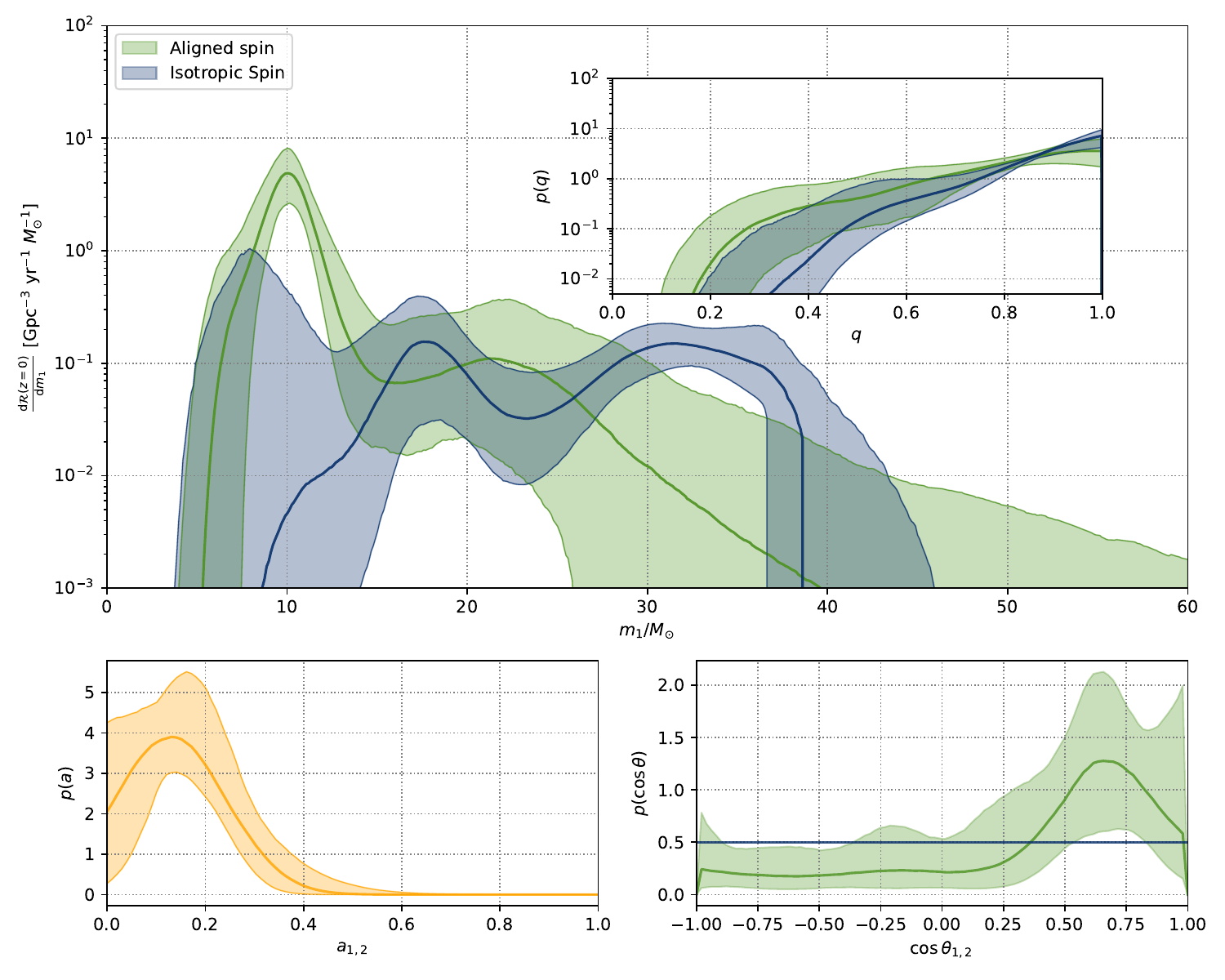}
\caption{The same as Figure~\ref{fig:Twopop_dist}, but for the inference with non-parametric $\cos\theta$-distribution in $[-1,1]$. The mass distributions of the two subpopulations are similar to those inferred using the \textsc{Extend Default} model.}
\label{fig:mact_spline_full}
\end{figure}

\cite{2022A&A...668L...2V,2023PhRvD.108j3009G,2023ApJ...946...16E} find that the $\cos\theta$-distribution is not peaking at $1$, i.e, the half-Gaussian model may be not appropriate for the $\cos\theta$-distribution distribution. Therefore, we modify the \textsc{Extend Default} model, where the half-Gaussian distribution is replaced with a non-parametric model $\mathcal{S}(\cos\theta) \propto e^{f(\cos\theta)}[z_{\rm min},1]$ \citep{2023PhRvD.108j3009G}. $f(x)$ is a cubic spline function defined by a set of nodes linearly located in $[z_{\rm min}, 1]$. Two cases are taken into consideration. The one is 4 nodes in $[0,1]$, which only allow the events with tilt angles $<90^{\circ}$, and the other is 6 nodes in $[-1,1]$. The prior on the amplitude of each node is a unit Gaussian distribution \citep{2023PhRvD.108j3009G}. Figure~\ref{fig:mact_spline} and Figure~\ref{fig:mact_spline_full} show the results of the two cases. Though the $\cos\theta$-distributions do not show peaks at $1$, there are bumps at $\cos\theta>0.5$ in both cases. The mass and spin-magnitude distributions of the two subpopulations are similar to Figure~\ref{fig:restrict} and Figure~\ref{fig:Twopop_dist} (inferred with parametric $\cos\theta$-distribution models).
We note that, if the nearly aligned subpopulation only allow the events with tilt angles $<90^{\circ}$, then the $10M_{\odot}$ peak in the primary-mass function is partially attributed to isotropic-spin subpopulation. In other words, some events in the $10M_{\odot}$ peak have tilt angles $>90^{\circ}$, which may be attributed to the dynamical formation channel in the globular clusters \citep{2024arXiv240114837T} or some special pathways producing BBHs with large misaligned spins in the isolated formation channel \citep{2021PhRvD.103f3032S}.


\subsection{Inference without selection effects}\label{app:nosel}

\begin{figure}
	\centering  
\includegraphics[width=0.8\linewidth]{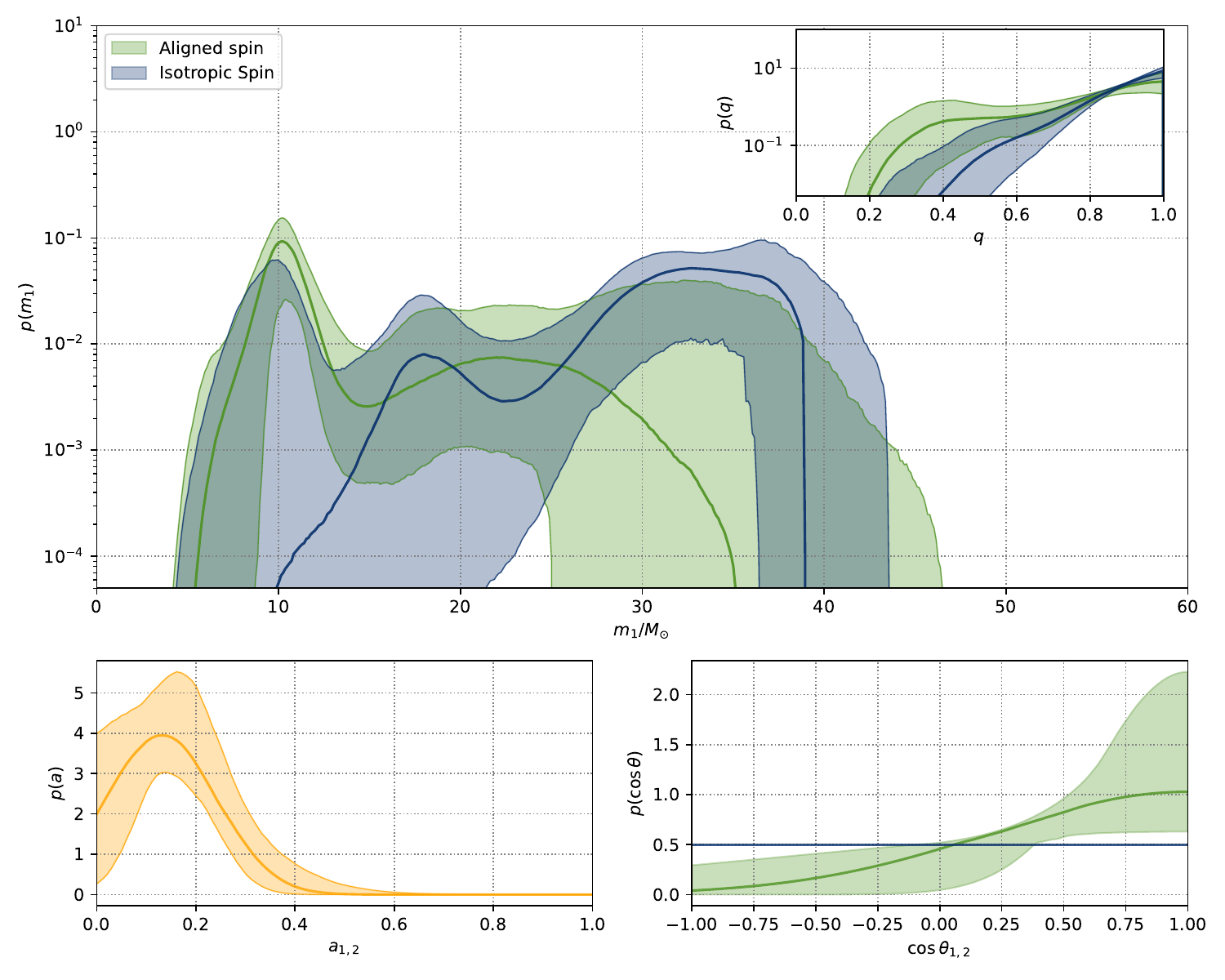}
\caption{The same as Figure~\ref{fig:Twopop_dist}, but for the inference without selection effects.}
\label{fig:nosel}
\end{figure}

We have also inferred without selection effects, as shown in Figure~\ref{fig:nosel}. We find the classification of the two subpopulations is similar to that inferred with mass-induced selection effects, though the slopes of mass functions are significantly changed. This indicates that the classification of the two subpopulations is insensitive to the mass-induced selection effects. 

\subsection{Extend Default model with the full catalog}\label{two_full}

\begin{figure}
	\centering  
\includegraphics[width=0.8\linewidth]{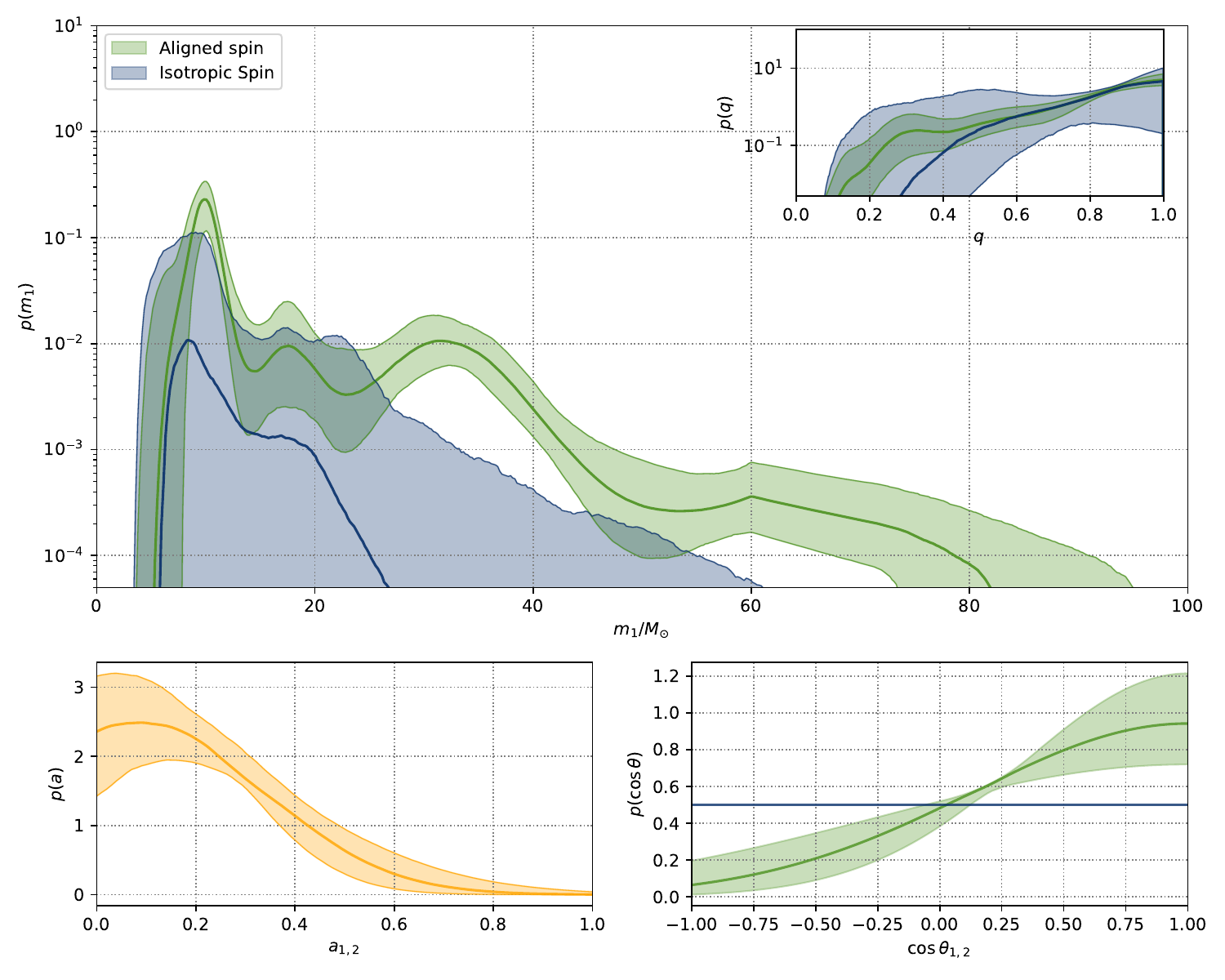}
\caption{The same as Figure~\ref{fig:Twopop_dist}, but for the inference with the full catalog.}
\label{fig:two_full}
\end{figure}

\begin{figure}
	\centering  
\includegraphics[width=0.8\linewidth]{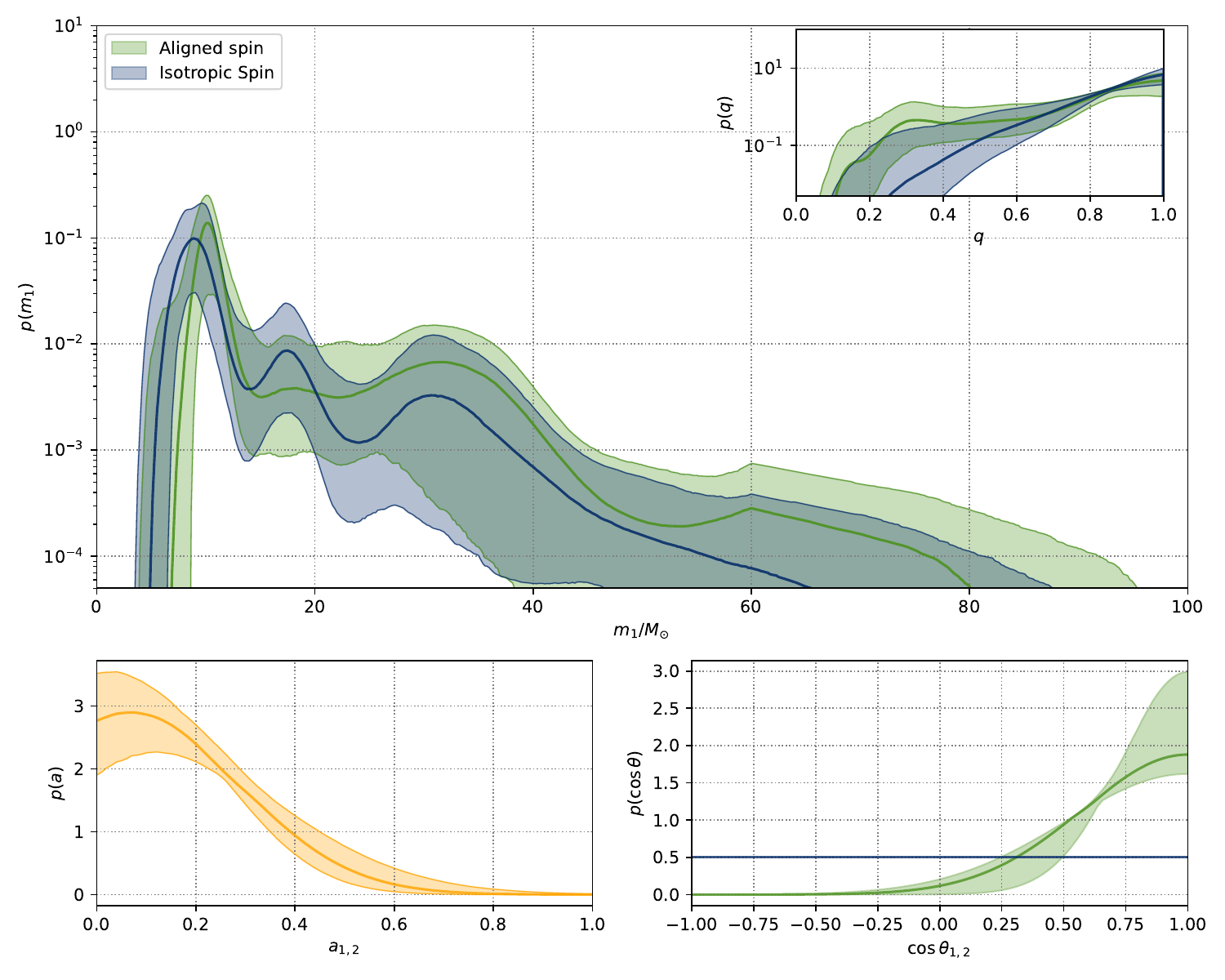}
\caption{The same as Figure~\ref{fig:two_full}, but for the inference with constraint $\sigma_{\rm t}<0.5$.}
\label{fig:two_full_rest}
\end{figure}

We have inferred using the \textsc{Extend Default} model with the full catalog, as shown in Figure~\ref{fig:two_full}. We found the results are significantly changed. However the \textsc{Extend Default} model is decisively disfavored compared to the \textsc{Main Model} by a Bayes factor of $\ln\mathcal{B}=7$. Additionally, we have also performed inference with $\sigma_{\rm t}<0.5$, and found the \textsc{Extend Default} model is even disfavored by a Bayes factor of $\ln\mathcal{B}=12$ (compared to the \textsc{Main Model} with $\sigma_{\rm t}<0.5$, the two subpopulations are degenerated with each other (see Figure~\ref{fig:two_full_rest}).
Considering the Bayes factors and the significant difference between the results of the \textsc{Extend Default} model and the \textsc{Main Model} fitting for the full catalog,
It is important to firstly model the high-spin subpopulation, when searching for the subpopulations via the spin orientations and the mass ratios.

\section{Classification}

Table~\ref{tab:psub} provides the probabilities of each first-generation event belonging to the field and dynamical channels. We find that the first detected BBH GW150914 \citep{2016PhRvL.116f1102A} originates from dynamical formation at $97.2\%$ credible level; while the popular asymmetric system GW190412 \citep{2020PhRvD.102d3015A} is more likely to originate from field binary evolution, which was examined to be promised by \cite{2020ApJ...901L..39O}. 
Additionally, the precession events GW190413\_134308, GW200129\_065458, and GW190521\_074359 \citep{2023arXiv230914473I} are more likely originates from dynamical channel.

\begin{longtable}{lcc}
\caption{Probabilities of the first-generation BBHs belonging to the field and dynamical subpopulations, inferred with the \textsc{Extend Default} model.\label{tab:psub}}
\\
\toprule
\hline
Events   &  Aligned & Isotropic \\ 
\midrule
\endfirsthead

\multicolumn{3}{c}{\autoref{tab:psub} ({\it Continued})}\\
\toprule
\hline
Events   &   Aligned & Isotropic \\ 
\midrule
\endhead
\bottomrule
\\
\multicolumn{3}{c}{\autoref{tab:psub} ({\it Continued on next page})}\\
\endfoot
\bottomrule
\\
\endlastfoot

GW150914\_095045&0.028&0.972 \\
GW151012\_095443&0.502&0.498 \\
GW151226\_033853&0.983&0.017 \\
GW170104\_101158&0.185&0.815 \\
GW170608\_020116&0.942&0.058 \\
GW170809\_082821&0.091&0.909 \\
GW170814\_103043&0.124&0.876 \\
GW170818\_022509&0.041&0.959 \\
GW170823\_131358&0.041&0.959 \\
GW190408\_181802&0.264&0.736 \\
GW190412\_053044&0.976&0.024 \\
GW190413\_134308&0.033&0.967 \\
GW190421\_213856&0.029&0.971 \\
GW190503\_185404&0.031&0.969 \\
GW190512\_180714&0.673&0.327 \\
GW190513\_205428&0.203&0.797 \\
GW190521\_074359&0.043&0.957 \\
GW190527\_092055&0.174&0.826 \\
GW190630\_185205&0.088&0.912 \\
GW190707\_093326&0.93&0.07 \\
GW190708\_232457&0.802&0.198 \\
GW190720\_000836&0.983&0.017 \\
GW190727\_060333&0.041&0.959 \\
GW190728\_064510&0.981&0.019 \\
GW190803\_022701&0.045&0.955 \\
GW190828\_063405&0.113&0.887 \\
GW190828\_065509&0.888&0.112 \\
GW190910\_112807&0.036&0.964 \\
GW190915\_235702&0.086&0.914 \\
GW190924\_021846&0.808&0.192 \\
GW190925\_232845&0.256&0.744 \\
GW190930\_133541&0.973&0.027 \\
GW190413\_052954&0.08&0.92 \\
GW190719\_215514&0.225&0.775 \\
GW190725\_174728&0.905&0.095 \\
GW190731\_140936&0.052&0.948 \\
GW191105\_143521&0.914&0.086 \\
GW191127\_050227&0.087&0.913 \\
GW191129\_134029&0.914&0.086 \\
GW191204\_171526&0.984&0.016 \\
GW191215\_223052&0.265&0.735 \\
GW191216\_213338&0.979&0.021 \\
GW191222\_033537&0.035&0.965 \\
GW200112\_155838&0.041&0.959 \\
GW200128\_022011&0.043&0.957 \\
GW200129\_065458&0.05&0.95 \\
GW200202\_154313&0.917&0.083 \\
GW200208\_130117&0.035&0.965 \\
GW200209\_085452&0.06&0.94 \\
GW200219\_094415&0.039&0.961 \\
GW200224\_222234&0.039&0.961 \\
GW200225\_060421&0.185&0.815 \\
GW200302\_015811&0.135&0.865 \\
GW200311\_115853&0.036&0.964 \\
GW200316\_215756&0.975&0.025 \\
GW191103\_012549&0.982&0.018 \\
GW200216\_220804&0.041&0.959 \\
\end{longtable}

\section{Simulation of mock data}\label{app:sim}

For the mock population with features found in this work, we assume the mass, spin, redshift distributions following Eq.~\ref{eq:sim} with $\alpha_{\rm A}=3.5$, $\beta_{\rm A}=1$, $\alpha_{\rm I}=0.5$, $\beta_{\rm I}=3$, $\delta^{\rm low}_{\rm A}=\delta^{\rm low}_{\rm I}=5M_{\odot}$, $m_{\rm min, A}=m_{\rm min, I}=5M_{\odot}$, $m_{\rm min, A}=m_{\rm min, I}=45M_{\odot}$, $\delta^{\rm high}_{\rm A}=\delta^{\rm high}_{\rm I}=10M_{\odot}$, $\sigma_{\rm t}=0.3$, $\mu_{\rm a}=0.15$, $\sigma_{\rm a}=0.1$, $r_{\rm I}=0.14$.
\begin{equation}\label{eq:sim}
\begin{aligned}
\pi(\boldsymbol{\theta} | {\bf \Lambda}) = [\pi_{\rm A}(\boldsymbol{\theta}_{\rm d} | {\bf \Lambda}_{\rm A})\times(1-r_{\rm I})+ \pi_{\rm I}(\boldsymbol{\theta}_{\rm d} | {\bf \Lambda}_{\rm I})\times r_{\rm I}]\times \pi_{\rm AI}(\boldsymbol{\theta}_{\rm s} | {\bf \Lambda}_{\rm AI})\\
= [\mathcal{PL}(m_1,m_2 |\alpha_{\rm A},\beta_{\rm A},\delta_{\rm A},m_{\rm min,A},m_{\rm max,A},\delta_{\rm A}^{\rm high})\times & \mathcal{G}(\cos\theta_1,\cos\theta_2|1,\sigma_{\rm t},-1,1)\times (1-r_{\rm I}) \\
+ \mathcal{PL}(m_1,m_2|\alpha_{\rm I},\beta_{\rm I},\delta_{\rm I},m_{\rm min,I},m_{\rm max,I},\delta_{\rm I}^{\rm high})\times & \mathcal{U}(\cos\theta_1,\cos\theta_2|-1,1)\times r_{\rm I}]\\
\times\mathcal{G}(a_1,a_2|\mu_{\rm a},\sigma_{\rm a},0,1)\times p(z|\gamma=2.7)
\end{aligned}
\end{equation}
the mass function of each subpopulation is 
\begin{equation}
\mathcal{PL}(m_1,m_2 |{\bf \Lambda_{\rm m}}) = C({\bf \Lambda_{\rm m}}) m_1^\alpha~S(m_1|\delta^{\rm low},m_{\rm min},m_{\rm max},\delta^{\rm high}) ~m_2^\alpha ~S(m_2|\delta^{\rm low},m_{\rm min},m_{\rm max},\delta^{\rm high})~(m_2/m_1)^\beta~\Theta(m_1-m_2).
\end{equation}
with 
\begin{equation}
S(m|\delta^{\rm low},m_{\rm min},m_{\rm max},\delta^{\rm high}) = SL(m|m_{\rm min},\delta^{\rm low})SH(m|m_{\rm max},\delta^{\rm high})
\end{equation}
where $C({\bf \Lambda_{\rm m}})$ is the normalization factor, $\Theta(m_1-m_2)$ denotes the Heaviside step function ensuring $m_1>m_2$, $SL(m|m_{\rm min},\delta^{\rm low})$ is the smooth function on the lower edge as introduced in \cite{2023PhRvX..13a1048A}, and $SH(m|m_{\rm max},\delta^{\rm high})$ is the smooth function on the upper edge which reads
\begin{equation}
\left\{
	\begin{aligned}
	&0  &&(m>m_{\rm max}), \\
	&[f(m_{\rm max}-m,\delta^{\rm high})+1]^{-1}  &&(m_{\rm max}-\delta^{\rm high}<m<m_{\rm max}),\\
	&1  &&(m<m_{\rm max}-\delta^{\rm high})
	\end{aligned}
	\right
	.
\end{equation}
with 
\begin{equation}
f(x,\delta^{\rm high}) = \exp({\delta^{\rm high}}/{x}+{\delta^{\rm high}}/({x-\delta^{\rm high}}))
\end{equation}

For the mock population without features found in this work, we just assume the $\cos\theta$ of the two subpopulation following the same distribution as $\mathcal{GU}(\cos\theta_1,\cos\theta_2|\sigma_{\rm t},\zeta)=\mathcal{G}(\cos\theta_1,\cos\theta_2|1,\sigma_{\rm t},-1,1)\times \zeta+ \mathcal{U}(\cos\theta_1,\cos\theta_2|-1,1)\times (1-\zeta)$, (i.e., the \textsc{Default} spin-orientation model in \cite{2023PhRvX..13a1048A},) with $\sigma_{\rm t}=0.3$ and $\zeta=0.5$. Therefore the mass, spin, redshift distributions read
\begin{equation}\label{eq:sim_no}
\begin{aligned}
\pi(\boldsymbol{\theta} | {\bf \Lambda}) = [\mathcal{PL}(m_1,m_2 |\alpha_{\rm A},\beta_{\rm A},\delta_{\rm A},m_{\rm min,A},m_{\rm max,A},\delta_{\rm A}^{\rm high})\times (1-r_{\rm I}) \\
+ \mathcal{PL}(m_1,m_2|\alpha_{\rm I},\beta_{\rm I},\delta_{\rm I},m_{\rm min,I},m_{\rm max,I},\delta_{\rm I}^{\rm high})\times r_{\rm I}]\\
\times\mathcal{GU}(\cos\theta_1,\cos\theta_2|\sigma_{\rm t},\zeta)\times\mathcal{G}(a_1,a_2|\mu_{\rm a},\sigma_{\rm a},0,1)\times p(z|\gamma=2.7)
\end{aligned}
\end{equation}
The mass and spin-orientation distributions are the same as those of the mock population with features, i.e., $\alpha_{\rm A},\beta_{\rm A},\delta_{\rm A},m_{\rm min,A},m_{\rm max,A},\delta_{\rm A}^{\rm high}, \alpha_{\rm I},\beta_{\rm I},\delta_{\rm I},m_{\rm min,I},m_{\rm max,I},\delta_{\rm I}^{\rm high},\mu_{\rm a},\sigma_{\rm a}$ are set the same as those in Eq.~\ref{eq:sim}.

To generate the detected events of the mock population we randomly choose events from the \href{https://zenodo.org/records/7890437}{injection campaign for O3 Search Sensitivity Estimates} with the inverse False Alarm Rate $>1 {\rm yr}$, each event is assigned a draw weight proportional to $p(\boldsymbol{\theta} | {\bf \Lambda})/p_{\rm draw}(\boldsymbol{\theta})$, where $p(\boldsymbol{\theta} | {\bf \Lambda})$ is the probability distribution of mock population, and $p_{\rm draw}(\boldsymbol{\theta})$ is the probability distribution from which the injection campaigns are drawn. For each mock population, we adopt 57 detections, similar to the size of the first-generation BBHs in the real data of GWTC-3. With each mock detection, we then use IMRPhenomXPHM waveform \citep{2021PhRvD.103j4056P} to generate GW signal and inject it to the noise generated by the \href{https://dcc.ligo.org/LIGO-T2000012/public}{“O3 actual” noise power spectral densities}. 
We perform parameter estimation on each event using BILBY \citep{2019ascl.soft01011A} with the NESSAI \citep{2021PhRvD.103j3006W} nested sampler.

\begin{figure}
	\centering  
\includegraphics[width=0.8\linewidth]{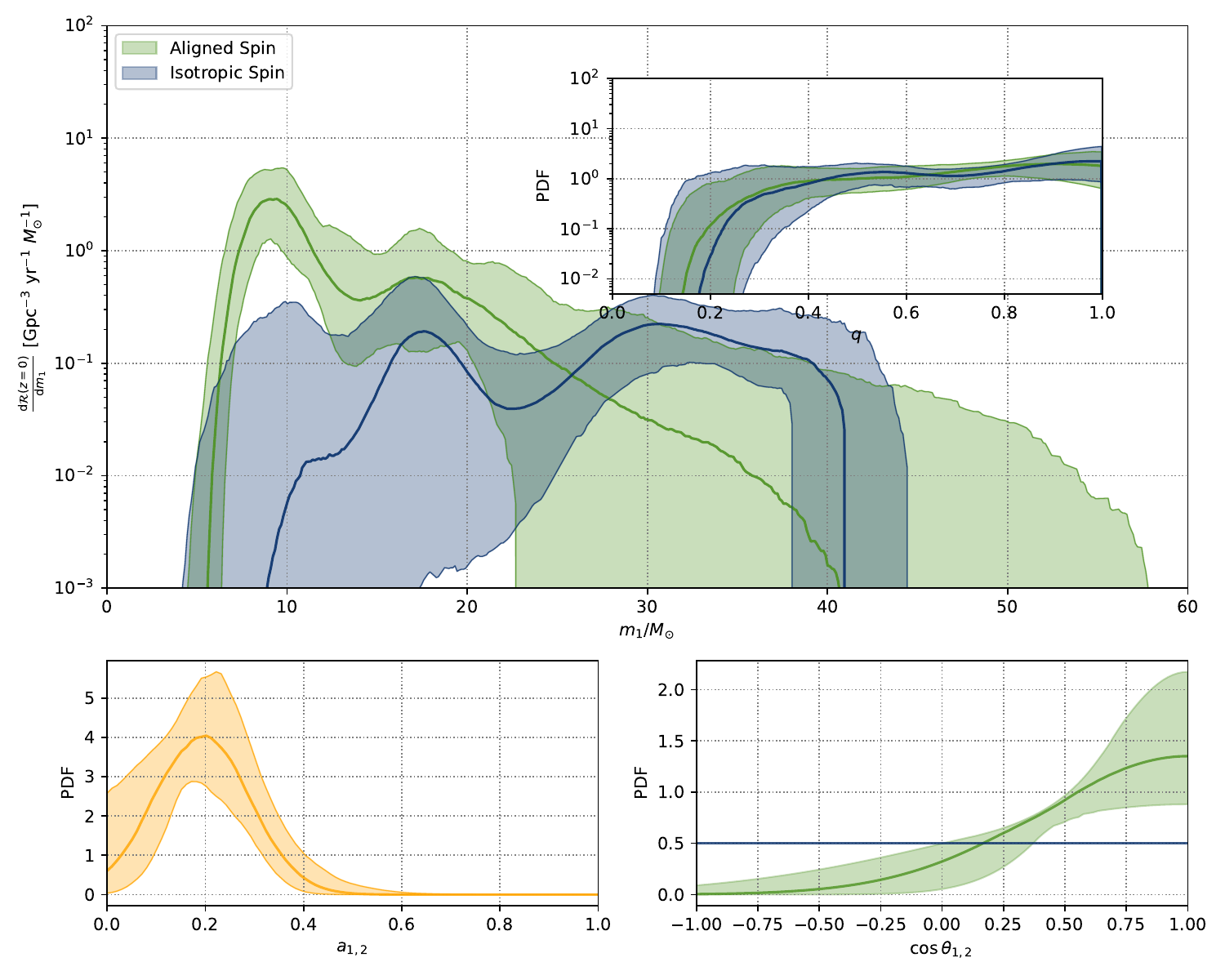}
\caption{The same as Fig.~\ref{fig:Twopop_dist}, but for the analysis with mock data where the injections having features found in this work.}
\label{fig:Withcorr_sim}
\end{figure}

\begin{figure}
	\centering  
\includegraphics[width=0.8\linewidth]{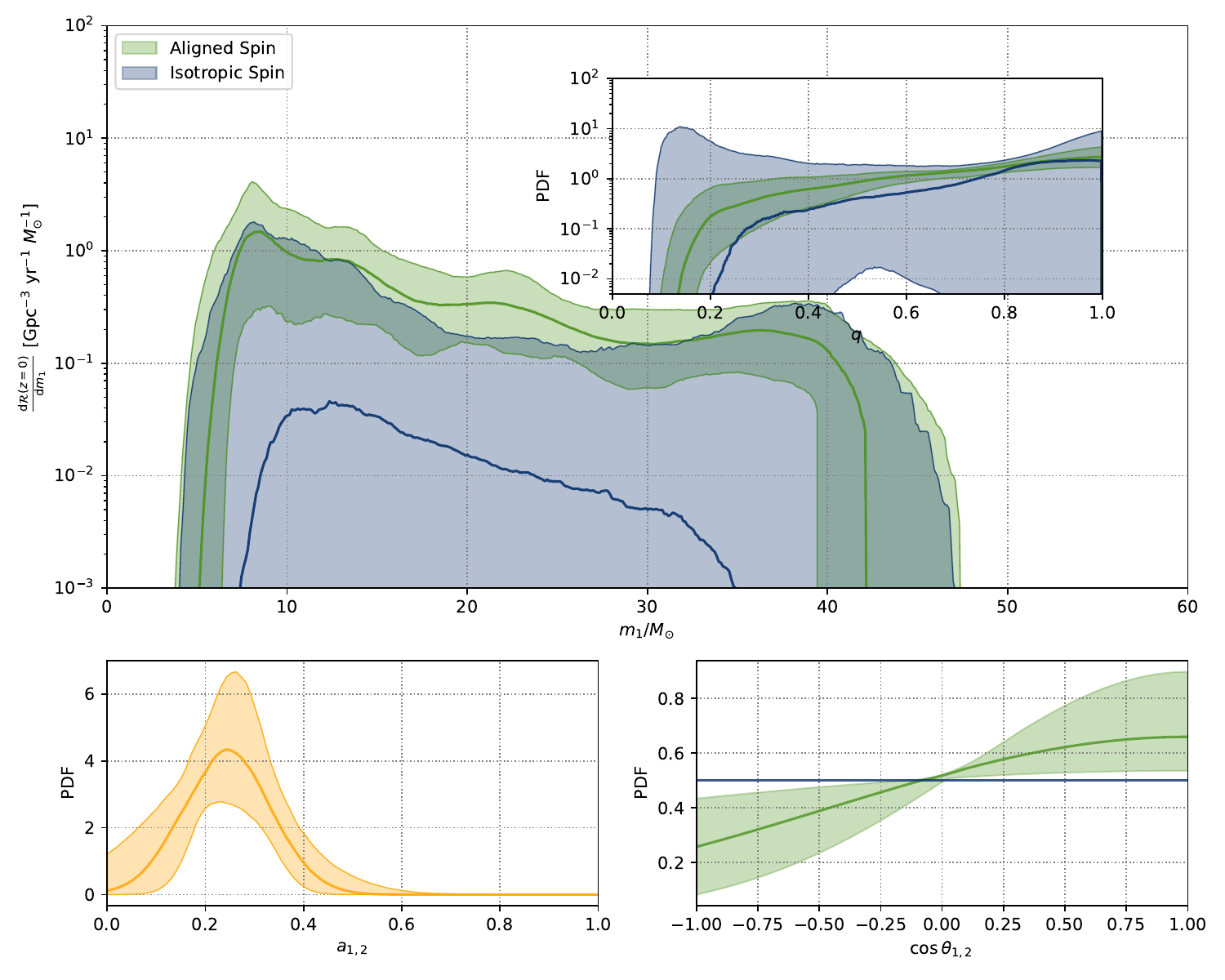}
\caption{The same as Fig.~\ref{fig:Withcorr_sim}, but for injections do not have features found in this work.}
\label{fig:Nocorr_sim}
\end{figure}

Then we use the model (Eq.~\ref{eq:twopop}) introduced in the main text, to respectively infer the underlying distribution of the two mock populations.
Figure~\ref{fig:Nocorr_sim} / Figure~\ref{fig:Withcorr_sim} shows the recovered distributions of the mock population with / without the features of spin-orientation distributions found in this work.

\bibliographystyle{aasjournal}
\bibliography{export-bibtex}

\end{CJK*}
\end{document}